\documentclass[journal,twoside]{IEEEtran}
\usepackage{subfigure}
\usepackage{graphicx}
\usepackage{amssymb}
\usepackage{amsmath}
\usepackage{cite}
\usepackage{multirow}
\usepackage{url}

\usepackage{algorithm}
\usepackage{algorithmicx}
\usepackage{algpseudocode}

\usepackage{pgfplots}
\pgfplotsset{compat=1.5}
\usepgfplotslibrary{groupplots}
\usepackage{hyperref}

\begin{document}

\title{Towards Hybrid-Optimization Video Coding}
\author{Shuai Huo,
Dong Liu, 
Li Li, 
Siwei Ma, 
Feng~Wu, 
and Wen Gao

%
\thanks{Date of current version \today. \emph{(Corresponding author: Dong Liu.)}
	
S. Huo, D. Liu, Li Li, and F. Wu are with the CAS Key Laboratory of Technology in Geo-Spatial Information Processing and Application System, University of Science and Technology of China, Hefei 230027, China (e-mail: huoshuai@mail.ustc.edu.cn; dongeliu@ustc.edu.cn; lil1@ustc.edu.cn; fengwu@ustc.edu.cn).

S. Ma and W. Gao are with the Institute of Digital Media, Peking University, Beijing 100871, China (email: swma@pku.edu.cn; wgao@pku.edu.cn).
%
}

}


\maketitle

\begin{abstract}
Video coding is a mathematical optimization problem of rate and distortion essentially.
To solve this complex optimization problem in practice, two popular video coding frameworks have been developed: block-based hybrid video coding and end-to-end learned video coding. If we rethink video coding from the perspective of optimization, we find that the existing two frameworks represent two directions of optimization solutions.
Block-based hybrid video coding represents the discrete optimization solution because those irrelevant coding modes are discrete in mathematics. The discrete solution provides multiple starting points (i.e. modes) in global optimization space and then searches for the best one among them. However, the search-based optimization algorithm is not efficient enough.
On the other hand, end-to-end learned video coding represents the continuous optimization solution because the optimization algorithm of deep learning, gradient descent, is based on a continuous function. The continuous solution optimizes a group of model parameters efficiently by such a numerical algorithm. However, limited by only one starting point, it is easy to fall into the local optimum.
To better solve the optimization problem, we propose a hybrid of discrete and continuous optimization video coding. We regard video coding as a hybrid of the discrete and continuous optimization problem, and use both search and numerical algorithm to solve it. Our idea is to provide multiple discrete starting points in the global space and optimize the local optimum around each point by numerical algorithm efficiently. Finally, we search for the global optimum among those local optimums.
Guided by the hybrid optimization idea, we design a hybrid optimization video coding framework, which is built on continuous deep networks entirely and also contains some discrete modes.
We conduct a comprehensive set of experiments to verify the efficiency of our hybrid optimization.
Compared to the continuous optimization framework, our method outperforms pure learned video coding methods. Meanwhile, compared to the discrete optimization framework, our method achieves comparable performance to HEVC reference software HM16.10 in PSNR.
\end{abstract}




\section{Introduction}
\label{sec_introduction}
Video coding is the enabling fundamental for most video technologies, including computer vision, image processing, visual communication, etc. Since the early days of digital video technology in the 1960s\cite{sikora2005trends}, academia and industry have started to study video coding and developed an outstanding framework called block-based hybrid video coding framework.
The name of block-based hybrid video coding is because it's a hybrid of two coding choices or modes (i.e. motion-handling and picture-coding techniques) at the block level. Under the two main choices, the framework allows to add various detailed modes, such as sub-pixel motion mode and merge mode in the motion-handling choice, skip and DCT mode in the picture-coding choice. Even if many modes are not very related to each other, this framework can still integrate them well and then searches for the best coding mode among them for each sequence. It is precisely because of the mode expansibility that an important development direction of video coding is to add more advanced modes based on the latest framework. Accordingly, a series of coding standards have been developed in the past forty years, including H.261, H.263, H.264, H.265, H.266\cite{bross2021developments}, EVC\cite{choi2020overview}, LCEVC\cite{meardi2020mpeg},  AV1\cite{han2021technical}, AVS3\cite{zhang2019recent}, etc. Now, H.266/VVC published by ITU-T and ISO/IEC in July 2020 is the state-of-the-art video coding standard.
However, with the increase of modes, the rate-distortion optimization by mode search becomes more complex.

Deep learning\cite{lecun2015deep} has brought about great breakthroughs and revolutionized the paradigm of image/video processing since 2012\cite{krizhevsky2012imagenet}.
Deep learning uses deep neural networks to convert the data into a representation at a more abstract level, and those representations are not hand-designed but learned from massive data.
With the help of the gradient descent algorithm, the deep networks can be optimized efficiently.
Inspired by that success, deep learning for image/video coding has also been explored since 2015\cite{dong2015compression,toderici2015variable}. Those works can be divided into two categories: deep tools and end-to-end methods. The deep tool is to add a deep network-based coding tool into the traditional coding framework. Those deep tools have improved the coding efficiency significantly, which confirmed the feasibility and potential of deep learning-based video coding. On the other hand, the end-to-end method is built entirely upon deep networks, and the whole framework optimizes a group of model parameters by gradient descent in an end-to-end manner. Although end-to-end works have more potential, due to the complexity of video coding tasks, the existing end-to-end works still have a little gap from the best conventional coding. Thus, the other key development direction of video coding is to design more effective network modules under the current end-to-end learned coding framework to realize its potential.

Video coding is to optimize the encoder-decoder parameters to achieve the best tradeoff performance of bit rate and distortion. Therefore, video coding is a mathematical optimization problem of rate and distortion essentially. If we look at video coding from the perspective of optimization, we find that the existing two frameworks represent two solution directions. The block-based hybrid video coding represents the discrete optimization solution. The various coding modes usually are not relevant to each other, so they are discrete in mathematics. Thus, this framework can be seen as providing multiple discrete points (i.e. coding modes) in the coding optimization space and then globally searching for the best mode for each coding sequence. Accordingly, the more discrete starting points will lead to better optimization results.
However, the search usually is not efficient enough for so many modes.
On the other hand, deep learning-based video coding represents the continuous optimization solution. In mathematics, gradient descent is based on continuous function.
Thus, deep learning-based video coding uses deep networks to formulate coding as a continuous optimization problem. Through the numerical algorithm (e.g. gradient descent), it optimizes a group of encoder-decoder parameters efficiently in continuous space. However, limited by the numerical algorithm and only one starting point, it is easy to fall into the local optimum.

We would like to ask whether discrete and continuous optimization can be combined together. The continuous method can optimize more efficiently by numerical algorithms, but continuous optimization usually provides only one starting point. On the other hand, discrete optimization can provide multiple starting points in the global space. Therefore, we are going a step further to ask whether we can provide multiple discrete starting points for continuous optimization to solve the video coding optimization problem. Based on this idea, we propose a hybrid of discrete and continuous optimization video coding. By means of discrete optimization, we provide various starting points distributed in the global optimization space. Through continuous optimization, we can optimize the local optimum around each starting point efficiently. Then we use the discrete search among those local optimums to obtain the global optimum. In theory, hybrid optimization is not only efficient but also more possible to achieve the globally optimal solution.

In this paper, we first analyze and summarize existing video coding works from the perspective of optimization.
We conclude that existing video coding frameworks adopt discrete or continuous optimization essentially, and we analyze why they will lead to the coding gain theoretically.
Inspired by the theoretical analysis, we propose a hybrid optimization video coding theory, and then design a hybrid optimization video coding framework. Our framework is built on deep networks entirely to achieve continuous optimization, and we add multiple starting points (i.e. multiple modes) to achieve discrete optimization.
We conduct a comprehensive set of experiments to verify the efficiency of our hybrid optimization.
Experimental results show that compared to the continuous optimization framework, our proposed method outperforms the pure end-to-end learned video coding methods. Meanwhile, compared to discrete optimization framework, our proposed method achieves comparable performance to HEVC reference software HM16.10 in PSNR.

The remainder of this paper is organized as follows. In Section \ref{sec_review}, we review the existing video coding works from the perspective of optimization.
In Section \ref{sec_Theoretical Analysis}, we theoretically analyze the optimization in video coding and then proposes a hybrid-optimization theory.
Section \ref{sec_Proposed Method} presents our designed hybrid-optimization coding framework guided by that theory.
Section \ref{sec_Implementation} presents the implementation details and Section \ref{sec_Optimization Procedure} presents the hybrid optimization procedure.
In Section \ref{sec_Experimental_Results}, we show the experimental results.
Section \ref{sec_Discussion} discuss the benefits and potential of proposed framework, and Section \ref{sec_Conclusion} concludes this paper.

\begin{table*}[]
\renewcommand\arraystretch{1.3}
\centering
\caption{Review of Optimization in Video Coding}
\begin{tabular}{c|c|c|c|c|c}
\hline
Category    & \begin{tabular}[c]{@{}c@{}}Optimization\\Space Formulation\end{tabular}    & \begin{tabular}[c]{@{}c@{}}Optimization\\ Method\\ (Offline)\end{tabular} & \begin{tabular}[c]{@{}c@{}}Optimization\\ Method\\ (Online)\end{tabular} & \begin{tabular}[c]{@{}c@{}}Optimization\\ Objective\end{tabular}   & Representative Works  \\ \hline
\begin{tabular}[c]{@{}c@{}}Classic Block-based Hybrid\\ Video Coding\\ (1984-now) \end{tabular} & Discrete    & N/A                         & Search                     & Approximate                  & \begin{tabular}[c]{@{}c@{}}H.120, H.261, MPEG-1, MPEG-2,\\H.263, H.264, H.265\end{tabular}     \\ \hline
\multirow{4}{*}{\begin{tabular}[c]{@{}c@{}}Block-based Hybrid Video Coding\\ with\\ Online Numerical Optimization\\ (1996-now)\end{tabular}}
& \multirow{2}{*}{Discrete \&} & \multirow{2}{*}{N/A} & \multirow{2}{*}{Search} & \multirow{4}{*}{Approximate} & \multirow{4}{*}{\begin{tabular}[c]{@{}c@{}}H.266 (Affine ME, ALF, BIO,\\ CCLM), AIF, LM, LIC\end{tabular}} \\
&                    &                    &                    &                    &                    \\\cline{3-4}
& \multirow{2}{*}{Locally Continuous} & \multirow{2}{*}{N/A} & \multirow{2}{*}{Numerical} &                    &                    \\
&                    &                    &                    &                    &                    \\
\hline
\multirow{4}{*}{\begin{tabular}[c]{@{}c@{}}Block-based Hybrid Video Coding\\ with\\ Deep Tools\\ (2015-now)\end{tabular}}
& \multirow{2}{*}{Discrete \&} & \multirow{2}{*}{N/A} & \multirow{2}{*}{Search} & \multirow{4}{*}{Approximate} & \multirow{4}{*}{Deep Tools, H.266 (MIP)} \\
&                    &                    &                    &                    &                    \\\cline{3-4}
& \multirow{2}{*}{Locally Continuous} & \multirow{2}{*}{Numerical} & \multirow{2}{*}{N/A} &                    &                    \\
&                    &                    &                    &                    &                    \\
\hline
\begin{tabular}[c]{@{}c@{}}End-to-End Learned \\ Image/Video Coding\\ (2015-now) \end{tabular}      & Globally Continuous   & Numerical   & N/A   & End-to-End                   & End-to-End Learned Video Coding                         \\ \hline
\end{tabular}
\label{table Optimization_Review}
\end{table*}

\begin{table*}[]
\renewcommand\arraystretch{1.3}
\centering
\caption{Overview of Representative Works in Video Coding Optimization}
\begin{tabular}{cccl}
\hline
\textbf{Method}                                                        & \textbf{Year}          & \textbf{\begin{tabular}[c]{@{}c@{}}Optimization\\ Objective\end{tabular}} & \multicolumn{1}{c}{\textbf{Bisic idea}}                                                                    \\ \hline \hline
\multicolumn{4}{c}{\textbf{Classic Block-based Hybrid Video Coding}}                                                                                                                                                                                                                                                                                                                                 \\ \hline \hline
H.120                                                                  & 1984                                                                                                                    & D                                                            & \begin{tabular}[c]{@{}l@{}}Propose the CR scheme with 2-mode searching, and \\ introduce the multi-modes optimization idea for the first time\end{tabular}                                                                                \\ \hline
H.261                                                                  & 1991                                                                                                                    & D                                                            & \begin{tabular}[c]{@{}l@{}}Introduce the MCP mode with multiply MV choices, and build\\ the block-based hybrid video coding framework for the first time\end{tabular}                                                                         \\ \hline
\begin{tabular}[c]{@{}c@{}}MPEG-1, MPEG-2,\\ H.263\end{tabular}         & \begin{tabular}[c]{@{}c@{}}1993, 1994,\\ 1996\end{tabular}                                                             & D                                                            & \begin{tabular}[c]{@{}l@{}}Introduce more modes, such as half-pixel motion, variable block sizes, etc,\\but still consider only distortion in the optimization objective\end{tabular} \\ \hline
\begin{tabular}[c]{@{}c@{}}Rate-Distortion\\ Optimization\end{tabular} & 1998                                                                                                                    & RD                                                            & \begin{tabular}[c]{@{}l@{}}Consider both rate and distortion in the optimization objective\\ for the first time\end{tabular}  \\ \hline
H.264, H.265                                                            & 2003, 2012                                                                                                             & RD                                                            & \begin{tabular}[c]{@{}l@{}}Introduce more modes, such as quarter-pixel motion, merge, etc,\\and adopt RD optimization but use approximate RD cost in ME\end{tabular}                      \\ \hline
\hline
\multicolumn{4}{c}{\textbf{Block-based Hybrid Video Coding with Online Numerical Optimization}}                                                                                                                                                                                                                                                                                                                                                                                     \\ \hline \hline
Accelerate Classic ME                                                                 & 1996                                                                                                           & RD                                                           & Search for MV towards the direction of gradient descent                                                                                                        \\ \hline
Accelerate Affine ME                                                                  & 2018                                                                                                           & RD                                                           & \begin{tabular}[c]{@{}l@{}}Optimize affine parameters towards the direction of\\ gradient descent in a high dimensional continuous space\end{tabular}                                                                                                       \\ \hline
\begin{tabular}[c]{@{}c@{}}Linear Model\\ (Linear Prediction,\\ LIC, CCLM)\end{tabular} & \begin{tabular}[c]{@{}c@{}}1998,\\ 2015,\\ 2018\end{tabular}                                                 & RD                                                           & \begin{tabular}[c]{@{}l@{}}Optimize the linear model by least squares method to predict current pixel\\from the neighboring pixels, or the reference pixels, or another component pixels\end{tabular} \\ \hline
\begin{tabular}[c]{@{}c@{}}Adaptive\\ Interpolation Filter\end{tabular}               & 2003                                                                                                           & RD                                                           & \begin{tabular}[c]{@{}l@{}}Optimize the Wiener filter by least squares method to perform adaptive \\ sub-pixel interpolation on the reference frame\end{tabular}                                           \\ \hline
\begin{tabular}[c]{@{}c@{}}Adaptive\\ Loop Filter\end{tabular}                        & 2008                                                                                                           & RD                                                           & \begin{tabular}[c]{@{}l@{}}Optimize the Wiener filter by least squares method to perform adaptive\\ loop filtering on the current reconstructed frame\end{tabular}                                          \\ \hline
\begin{tabular}[c]{@{}c@{}}Bi-Directional\\ Optical Flow\end{tabular}                 & 2010                                                                                                           & RD                                                           & \begin{tabular}[c]{@{}l@{}}Calculate the analytic solution of optical flow gradient to refine the bi-prediction\end{tabular}
\\ \hline \hline
\multicolumn{4}{c}{\textbf{Block-based Hybrid Video Coding with Deep Tools}}                                                                                                                                                                                                                                                                                                                                                                                                          \\ \hline \hline
Filtering Tools                                                                       & 2015                                                                                                           & D                                                            & Optimize DNN to refine the reconstructed frame                                                                                                                              \\ \hline
Prediction Tools                                                                      & 2018                                                                                                           & RD                                                           & Optimize DNN to perform or refine the intra/inter prediction                                                                                                                \\ \hline
Transform Tools                                                                       & 2018                                                                                                           & RD                                                           & Optimize DNN to perform the transform                                                                                                                                       \\ \hline
\begin{tabular}[c]{@{}c@{}}Entropy Coding Tools\end{tabular}                        & 2017                                                                                                           & R                                                            & \begin{tabular}[c]{@{}l@{}}Optimize DNN to predict the probability distribution based on the context\end{tabular}                                                        \\ \hline
\begin{tabular}[c]{@{}c@{}}Down- and Up- \\ Sampling Tools\end{tabular}                & 2018                                                                                                           & RD                                                           & Optimize DNN to achieve the adaptive resolution coding                                                                                                                      \\ \hline
\begin{tabular}[c]{@{}c@{}}MIP\end{tabular}                                            & 2018                                                                                                           & RD                                                           & \begin{tabular}[c]{@{}l@{}}Optimize simplified DNN in H.266 to predict the current block from reference pixels.\\ This is the first deep tool adopted by video coding standards.\end{tabular}
\\ \hline
\begin{tabular}[c]{@{}c@{}}Accelerate Block\\Partition/Mode Decision\end{tabular}                       & 2016                                                                                                     & Accuracy                                                                  & \begin{tabular}[c]{@{}l@{}}Optimized DNN to decide the CU partition result directly\\or to reduce the mode searching range\end{tabular}
\\ \hline
\multicolumn{4}{c}{\textbf{End-to-End Learned Image/Video Coding}}                                                                                                                                                                                                                                                                                                                                                                                                                                  \\ \hline \hline
\begin{tabular}[c]{@{}c@{}}Variable-Rate RNN-\\ based Image Coding\end{tabular}             & 2015                                                                                                     & D                                                                         & \begin{tabular}[c]{@{}l@{}}End-to-end optimize DNN for image compression for the\\first time, but use only distortion in the optimization objective\end{tabular}                 \\ \hline
\begin{tabular}[c]{@{}c@{}}RD Optimization-\\ based Image Coding\end{tabular}               & 2017                                                                                                     & RD                                                                        & \begin{tabular}[c]{@{}l@{}}End-to-end optimize DNN for image compression by RD cost\end{tabular}  \\ \hline
\begin{tabular}[c]{@{}c@{}}Deep Video \\ Compression (DVC)\end{tabular}                     & 2019                                                                                                     & RD                                                                        & \begin{tabular}[c]{@{}l@{}}End-to-end optimize DNN for video compression for the first time\end{tabular}                                                                          \\ \hline
\begin{tabular}[c]{@{}c@{}}Hierarchical \\ Learned Video \\ Compression (HLVC)\end{tabular} & 2020                                                                                                     & RD                                                                        & \begin{tabular}[c]{@{}l@{}}End-to-end optimize DNN for B-frame compression with hierarchical quality layers\end{tabular}                                                                  \\ \hline
\begin{tabular}[c]{@{}c@{}}Feature Video \\ Compression (FVC)\end{tabular}                  & 2021                                                                                                     & RD                                                                        & \begin{tabular}[c]{@{}l@{}}End-to-end optimize DNN for video compression in the feature space\end{tabular}                                                                                \\ \hline
\begin{tabular}[c]{@{}c@{}}Online Encoder \\ Updating\end{tabular}                          & 2020                                                                                                     & RD                                                                        & \begin{tabular}[c]{@{}l@{}}Online end-to-end optimize the DNN-based encoder in deep video compression\\for each test sequence\end{tabular}                                                                  \\ \hline
\begin{tabular}[c]{@{}c@{}}Resolution-adaptive\\ Flow Coding\end{tabular}                   & 2020                                                                                                     & RD                                                                        & \begin{tabular}[c]{@{}l@{}}Online search from multi-resolution modes in deep video compression\end{tabular}
\\ \hline \hline
\end{tabular}
\label{table Representative Work_Review}
\end{table*}

\section{Review of Optimization in Video Coding}
\label{sec_review}
In this section, we review the development of rate-distortion optimization in video compression.
Theoretically, the optimization objective in video coding is: minimize distortion $D$, subject to a constraint $R_c$, on the number of bits used $R$.
To solve the optimization of video coding, it's necessary to give a coding framework or model (i.e mathematical function expression) firstly.
Based on the model, the solving process can generally be divided into two steps\cite{ortega1998rate}: (i) offline step: optimize initial parameters of the model from a large number of training sequences; (ii) online step: striving to further optimize parameters of the model for each test sequence to obtain the best coding performance.
Therefore, offline step focuses on how to get the general model parameters as initial optimization points, while online step concentrates on how to get the special model parameters for each coding sample as the final optimization solution.
Although all video coding optimization algorithms follow them, those algorithms have different emphasis on the two steps and then adopt different optimization ideas correspondingly.
According to those differences, the rate-distortion optimization in video compression can be divided into four categories.

The first is the optimization in classic block-based hybrid video coding, which is the earliest and most widely used optimization scheme. On the offline step, it directly sets the parameters of modes in the framework without optimization. Then it searches for the optimal mode online from some designed modes for a given coding sequence. When the number of modes is small, it is an effective optimization scheme. However, with the increase of modes, the searching cost becomes higher and higher, making the coding framework hard to be optimized.

To address this problem, the second kind of optimization scheme has appeared, which utilizes some numerical algorithms to online optimize some parts of block-based hybrid video coding more efficiently.
The first and second classes of optimization concentrate on the online step, so they can achieve adaptive coding for each sequence. Unfortunately, the encoders with too high calculation cost at the online step will seriously affect the practical application of coding. However, the computational cost at the offline step has less effect on the coding.

Accordingly, the third kind of optimization scheme has emerged in block-based hybrid video coding, which adopts the numerical optimization algorithms at the offline step to achieve more efficient optimization for some parts. Those algorithms usually train a deep network to replace a hand-optimized module of block-based hybrid video coding, so they are also called deep tools. Numerical optimization can not only speed up the entire optimization process but also obtain finer optimization results, so it shows great potential. However, the deep tools only use numerical optimization on some parts of the coding framework. In order to fully explore the potential of numerical optimization, it is natural to think whether all the search-based modules in the coding framework can be replaced with numerical optimization ones.

Then the fourth kind of optimization scheme was proposed, in which the whole coding framework adopts the numerical optimization algorithms at the offline step. Those frameworks are usually on top of deep networks. What's more, those frameworks are end-to-end trained at the offline step and directly infer the coding results for each sequence at the online step, so they are also called end-to-end learned video coding.

In the following, we analyze the four categories of rate-distortion optimization in video compression and corresponding coding technologies in detail. The characteristic of each category is summarized in Table \ref{table Optimization_Review}, and the representative works of each category are analyzed in detail in Table \ref{table Representative Work_Review}.

\subsection{Classic Block-based Hybrid Video Coding}
\label{sec_Classic Block-based Hybrid Video Coding}
Motion video data consists essentially of a time-ordered sequence of pictures. The most straightforward method of compressing video content is to compress each picture simply using an image-coding syntax.
This method uses only one image-coding mode to deal with all video content.
However, much of the depicted scene is essentially just repeated in picture after picture without any significant change. Image coding for video data doesn't take advantage of temporal redundancy.

A simple solution is coding only the changes in a video scene.
Based on this idea, the first international digital video coding standard H.120\cite{ITU-T1984H120} was developed by the ITU-T organization and received final approval in 1984.
H.120 offline designs a conditional replenishment (CR) method to selectively code the change. CR consists of two coding modes: SKIP and INTRA mode.
The SKIP mode is sending signals to indicate which areas are repeated, and the INTRA mode is sending new coded information to replace the changed areas.
Accordingly, the INTRA mode is used to handle the large scene changes, while the SKIP mode can effectively deal with the repeated scenes. CR allows searching the best choice between two modes of representation for each area online.
Although CR is very simple in the algorithm, its optimization ideas lay the foundation for later video coding standards.
That is, CR introduces multi-mode optimization, which designs various modes for different situations and searches for the best mode among those discrete modes.
At this time, the optimization space in CR is still small because it only has two modes.

CR coding only allows exact repetition or complete replacement, but the content of a prior picture often can be a good approximation.
Based on this idea, the first practical success video coding standard H.261 \cite{ITU-T1991H261} is approved by the ITU-T in early 1991, which can be capable of operation at affordable telecom bit rates.
Specifically, the motion estimation (ME) in the encoder searches for the best integer spatial displacement. The motion-compensated prediction (MCP) uses the spatial displacement of the prior picture to form an approximation prediction.
Next, the displaced frame difference (DFD) coding method codes the resulting difference to refine the MCP signal.
Meanwhile, the processing unit is based on the block ($16\times16$ macroblock in MCP and $8\times8$ block in DCT).
Accordingly, the basic coding structure in H.261 is called block-based hybrid video coding due to its construction as a hybrid of motion-handling and picture-coding techniques at the block level.
H.261 was the first standard to use the basic typical structure we find still predominant today.
Compared to H.120, H.261 still uses multi-mode optimization but greatly enlarges the optimization space. Because in addition to CR mode choices, the ME and MCP modes provide a large amount of discrete motion vector (MV) choices.
With the help of more mode choices, the representative ability of the offline designed model improves significantly.
Then the better coding choice is searched for online among those choices, so the coding performance is improved.

Based on the H.261 hybrid coding framework and multi-mode optimization, a series of standards were approved subsequently, including MPEG-1 \cite{ISO1993MPEG-1} in 1993, MPEG-2 \cite{ITU-T/ISO1994MPEG-2} in 1994, and H.263 \cite{ITU-T1995H263} in early 1996.
They add more advanced modes, such as half-pixel motion mode and bi-prediction mode in MPEG-1, variable block-size motion compensation mode ($16\times16$ and $8\times8$) in H.263, and so on.
More modes expand the optimization space further, so the coding performance is improved further.
However, in all those standards, the multi-mode search criterion just considers distortion but not rate, so the optimization objective is approximate.

To address this problem, the rate-distortion optimization method \cite{sullivan1998rate,ortega1998rate} was proposed in 1998, which considers both the rate and distortion in the optimization objective. These works theoretically analyze that the video coding needs to optimize both the rate and distortion, and then they introduce the Lagrangian method to solve the optimization problem effectively.
The Lagrangian optimization changes the constrained optimization problem into an unconstrained problem so that the optimization objective becomes minimizing the Lagrangian rate-distortion cost $D + \lambda R$.
This rate-distortion optimization has a significant impact on video coding.
All later video coding standards adopt this RD optimization, such as H.264 \cite{wiegand2003overview}, H.265 \cite{sullivan2012overview}, and so on.
Those standards contain more modes and use RD optimization to search for the best one effectively.
In the specific implementation, the optimization objective in most modes is accurate, where $R$ and $D$ are based on the final stream and reconstructed frame, respectively.
However, in ME, due to the limitation of computational power, the optimization objective is approximated, i.e., $R$ and $D$ are based on the MV bits and prediction frame.
Therefore, the optimization objective in those standards is still approximate.

From the above review, we can find that the mentioned block-based hybrid video coding standards have some same characteristics in optimization.
First, all the offline designed models in those standards consist of many discrete modes\cite{wiegand2003rate,ohm2012comparison}, including INTRA mode, SKIP mode, partition mode, MCP mode with lots of MV choices, etc. Therefore, the optimization parameter space in those standards is discrete.
Second, at the offline step, the model parameters are almost not optimized but hand-crafted. It makes efforts to online search for the optimal mode from the discrete modes for each sequence. Thus, the optimization method is online searching.
Third, the optimization objective is approximate rather than actual RD cost.

Although this multi-mode search optimization has achieved great success, there are still some problems.
First, with the increase of modes, the cost of searching has increased significantly, making the framework hard to be optimized.
Second, search optimization can only conduct on a finite number of discrete modes.
Considering the searching complexity, it can't perform more refined optimization on the parameters.

\subsection{Block-based Hybrid Video Coding with Online Numerical Optimization}
\label{sec_Block-based Hybrid Video Coding with Online Numerical Optimization}
To address the inefficient problem of search optimization, some online numerical optimization methods appeared, which replace part of the search optimization in block-based hybrid video coding.
Compared to search optimization, numerical optimization \cite{nocedal2006numerical} not only is more efficient but also makes the optimization parameters more refined.
According to the two advantages, the existing numerical optimization methods can be divided into two categories. The first is for improving optimization speed, and the second is to increase compression efficiency.

The first is mainly to use gradient descent-based method to accelerate the multi-modes search.
In the classical block-based hybrid coding, it is difficult to establish a simple mathematical model to associate different characteristic modes, such as INTRA mode, SKIP mode, and MCP mode. However, the different MV choices within MCP mode and RD cost can be easily modeled with the mathematical function. Based on the characteristic, in 1996, an online gradient descent optimization algorithm for ME was proposed \cite{liu1996block}.
It calculates the gradient of the objective function to MV, and the search for MV always moves in the direction of optimal gradient descent.
Compared to the exhaustive search on discrete points, gradient-based optimization can fully use the continuous characteristics and exclude many unnecessary search points.
Accordingly, it can conduct ME more efficiently.
Later on, a series of gradient descent-based optimization for ME was developed \cite{Dufaux2000Efficient,chen2000motion,po2009novel}.
However, because the search points of classical MVs are not many enough, search-based optimization can still handle ME well. Thus, gradient descent-based optimization has not attracted widespread attention.

Until recent years, it's hard for the classical translational motion model to improve performance further, so the more advanced affine motion model \cite{li2018efficient} has been adopted in the latest video coding standards H.266/VVC \cite{bross2020versatile,bross2021overview}.
The affine motion parameters are no longer simple MVs, but more complex, larger range and higher precision control parameters, which is more difficult for the classical search-based ME.
To efficiently optimize the affine motion parameters, online gradient descent-based ME plays a key role.
The gradient-based method can not only handle the parameters in a high dimensional continuous space \cite{wang2002video}, such as affine parameters, but also quickly optimize the best parameters with only a few iterations \cite{li2018efficient}.
However, the optimization objective is still approximate like the classical ME.

The second category is to calculate the analytic solution by some numerical methods to perform more refined optimization, including least-squares optimization and others.
The least-squares optimization is to fit the mapping relationship with the smallest error for each group of signals.
It is mainly applied in adaptive prediction and filter in video coding.
As for the adaptive prediction, those works use least squares to online optimize the linear model (LM) on pixels within or across the component.
Those works build the linear model to predict the current pixel from the neighboring pixels (called linear prediction) \cite{wu1998piecewise,li2001edge,ye1999least}, or the reference pixels (called local illumination compensation (LIC)) \cite{Liu2015Local}, or another component pixels (called cross-component linear model (CCLM)) \cite{zhang2018enhanced}.

As for the filter, the work is to use least squares to optimize the Wiener filter to conduct adaptive filtering.
But the filter are operated on different frames, which is either on the reference frame \cite{wedi2003motion,wedi2006adaptive,vatis2008adaptive} (called adaptive interpolation filter (AIF)) or in the current reconstructed frame \cite{tsai2013adaptive,karczewicz2021vvc} (called adaptive loop filter (ALF)).
AIF uses some Wiener filters to interpolate multiply sub-pixel images adaptively from the reference frames, and then those frames are used by MC to generate a more accurate prediction.
ALF uses the Wiener filters to enhance the current reconstructed frame adaptively.
By efficiently and finely online optimizing the Wiener filter with the least squares method, AIF and ALF significantly improve coding performance. Meanwhile, ALF also is adopted by H.266 standard as a core tool.

In addition to those mentioned technologies, numerical methods also are applied in other technologies in H.266, such as the bi-directional optical flow (BDOF)\cite{alshin2010bi}. BDOF uses the optical flow concept to refine the bi-prediction signal. It uses numerical methods to online calculate the analytic solution of the optical flow gradient and then derives the refinement.
Compared to search optimization on some discrete points, numerical methods can online optimize the parameters of models (e.g. the linear model, adaptive filters, and optical flow model) in the continuous space to achieve more refined optimization.

From those works, we can find that the mentioned methods have some same characteristics in optimization.
First, these offline designed models are still based on discrete modes, that is, following block-based hybrid video coding, but they choose some highly correlated choices within a mode to make them locally continuous.
Second, the model optimization still focuses on the online step. It still uses the search-based method for discrete modes, while it uses the numerical method for continuous parts to achieve efficient and refined optimization.
Third, the optimization objective is approximate because those tools only considers the RD cost of their local parts rather than the whole RD cost, such as the prediction RD cost for gradient descent-based ME and prediction RD cost for AIF.

Although online numerical optimization is more efficient than search optimization, the optimization still focuses on the online step.
The high computational cost of online optimization will limit the wide application of the codec.

\subsection{Block-based Hybrid Video Coding with Deep Tools}
\label{sec_Block-based Hybrid Video Coding with Deep Tool}
Compared to online numerical optimization, the computational cost of offline optimization has less effect on the coding.
Correspondingly, some offline numerical optimization methods appeared.
It not only retains the two advantages of efficient and refined optimization in numerical optimization, but also changes the optimization stage from online to offline.
Those algorithms usually take offline trained deep networks as tools of block-based hybrid video coding, and then use the gradient descent-based numerical method to optimize the network.
Similarly, according to the two advantages of numerical optimization, the existing deep tools can also be categorized into two groups \cite{liu2020deep}. The first group is to increase compression efficiency, and the second is to improve encoding speed.

The first group of deep tools may either replace the corresponding hand-optimized module with a deep network, or newly add a deep network into the scheme.
It involves various tools, including prediction tools, transform tools, entropy coding tools, filtering tools, down- and up-sampling tools, and so on.
Since some online numerically optimized tools have been developed, the straightforward idea is to replace those online tools with offline optimized networks.
Among those tools, the filtering tool improves compression efficiency most significantly.
Correspondingly, the deep network with offline numerical optimization was first applied in the reconstruction filtering module in video coding. In 2015, Dong \emph{et al.} \cite{dong2015compression} propose a CNN for compression artifacts reduction on JPEG, namely ARCNN.
They offline designed the network with 4 convocation layers and optimized the parameters with the gradient descent method through lots of training samples.
In the test, ARCNN doesn't need online optimization, which directly infers the refined image.
Experimental results show that ARCNN achieves incredible performance, i.e. more than 1dB improvement in PSNR than JPEG.
ARCNN proved the feasibility of deep tools with offline numerical optimization and demonstrated its great potential.
Subsequently, a series of deep filtering tools are proposed \cite{dai2017convolutional,dai2018cnn,guan2019mfqe,jia2019content,zhang2018residual}, which occupies the majority of the deep tools.
Although the works have different network structures, their optimization ideas are the same as that of ARCNN.

Inspired by the great success of deep filtering tools, other deep tools have been proposed \cite{pfaff2018neural,zhao2018enhancedBi,liu2019one,ma2019iwave,ma2019convolutional,li2018fully,huo2018convolutional,Li2018,yan2018convolutional,yan2019invertibility,huo2021deep,liu2018cnn,song2017neural,li2018convolutional,li2019learning}.
Some of them replace other online numerically optimized tools with offline optimized networks, and some develop new tools.
For example, for the online linear prediction tool, Li \emph{et al.} \cite{li2018fully} propose an offline optimized fully connected network for intra prediction, which also predicts the current pixel from neighboring pixels.
Similar to the idea of LIC, Huo \emph{et al.} \cite{huo2018convolutional} propose an offline optimized CNN to refine the inter prediction signal.
Analogous to CCLM, Li \emph{et al.} \cite{Li2018} propose an offline optimized CNN to perform cross-channel prediction.
Like AIF, Yan \emph{et al.} \cite{yan2018convolutional,yan2019invertibility} propose to take offline optimized CNNs as filters to perform sub-pixel interpolation on the reference frames.
In addition to these existing tools with numerical optimization, there are also many newly developed offline optimized deep tools.
For example, for the prediction module, Huo \emph{et al.} \cite{huo2021deep} propose a CNN-based new inter prediction mechanism by extrapolating the current frame from the multiple reference frames and taking it as another reference frame.
For the transform module, Liu \emph{et al.} \cite{liu2018cnn} propose a CNN-based method to achieve a DCT-like transform for image coding.
For the entropy coding module, Song \emph{et al.} \cite{song2017neural} optimize a CNN to predict the probability distribution of the intra prediction mode based on the context.
For the down- and up-sampling coding module, Li \emph{et al.} \cite{li2018convolutional,li2019learning} propose a CNN-based down- and up-sampling method to achieve the adaptive resolution coding.

Due to the high compression efficiency, H.266 standard has adopted an offline numerical optimization-based tool, i.e. matrix-based intra-picture prediction (MIP) \cite{pfaff2018neural}.
MIP uses the neural network to predict the current block from the reference samples.
The neural network has been simplified to a matrix-vector multiplication, with the matrix being selected from a set of pre-trained matrices and the vector being constructed from the reference samples.
In order to offline optimize the matrix coefficients from lots of training samples, a gradient descent-based numerical algorithm (Adam \cite{kingma2014adam}) has been used.

The second group of deep tools is intended for improving encoding speed \cite{liu2016cu,xu2017reducing,jin2017cnn,kim2018fast,feng2021cnn}.
Based on the analysis in Section \ref{sec_Block-based Hybrid Video Coding with Online Numerical Optimization}, the choice within a specific mode and its RD cost can be easily modeled with the mathematical function.
Among the core coding modes, the MCP mode and the block partition mode have the most internal choices, i.e. multiply MV choices in MCP mode and various block combination choices in partition mode.
However, due to the complexity of motion and the massive MV choices, it is difficult for the offline optimized network to predict MV accurately.
Therefore, only a few offline numerical optimization methods accelerate MCP mode decision\cite{storch2021fastinter360}, and most methods focus on accelerating partition choice selection.
In 2016, Liu \emph{et al.} \cite{liu2016cu} presented an offline optimized CNN to help decide CU partition mode for HEVC intra encoder for the first time. Without searching among multiply partition choices, the trained CNN will directly decide whether to split CU or not based on the content and the specified QP.

From those works, we can find that the mentioned methods have some same characteristics in optimization.
First, these offline designed models are still based on discrete modes, i.e. following block-based hybrid video coding, but they make part modes or tools locally continuous.
Second, the model optimization is not only on the online step but also on the offline step.
The continuous parts are optimized offline by the numerical method, while the discrete parts are optimized online by searching.
Third, the optimization objective is approximate, because the optimization objective of those deep tools is based on either the accuracy (e.g. fast partition decision), or only D (e.g. ARCNN), or only R (e.g. CNN-based entropy coding), or part RD (e.g. CNN-based residual transform), rather than the actual RD cost.

Numerical optimization can not only speed up the optimization process but also perform more refined optimization.
However, the deep tools only use numerical optimization on some parts of the coding framework.
We would like to ask whether the whole coding framework can be optimized with numerical methods.

\subsection{End-to-End Learned Image/Video Coding}
\label{sec_End-to-End Learned Image/Video Coding}
Based on the idea of the whole coding framework with numerical optimization, end-to-end learned image/video coding methods were proposed.
The coding model is built on top of deep networks entirely.
At the offline step, one group of average optimal model parameters are end-to-end optimized by numerical algorithms through lots of training samples.
At the online step, the offline optimal model usually directly infers the coding results for each sequence, which is not optimized further.
Compared to deep tools with local numerical optimization, end-to-end learned coding can perform more efficient and more refined optimization for the whole framework.
Meanwhile, the whole framework takes the actual RD cost as the optimization objective and unites all numerical modules to optimize in an end-to-end manner.
Next, we will review the end-to-end learned image and video coding in turn.

Since image coding is simpler than video coding, the end-to-end learned method was initially tried in image coding.
In 2015, Toderici \emph{et al.} \cite{toderici2015variable} proposed an offline end-to-end optimized RNN-based framework for variable rate image compression.
They use binary quantization to generate codes, and do not consider the rate during offline training, i.e. the optimization objective is only end-to-end distortion.
This work builds a coding framework with all numerical optimization modules for the first time and initially verifies the feasibility.
However, the optimization objective only based on distortion limits the improvement of efficiency.

To address this problem, many rate-distortion optimized end-to-end image compression frameworks are proposed \cite{balle2016end,balle2018variational,minnen2018joint,toderici2017full,johnston2018improved,ma2020end,hu2021learning,cheng2020learned}.
The most representative works of CNN-based methods are from Ball{\'e} \emph{et al.} \cite{balle2016end,balle2018variational,minnen2018joint}.
Ball{\'e} \emph{et al.} proposed various end-to-end RD optimized image coding frameworks, using the factorized-prior\cite{balle2016end}, hyper-prior \cite{balle2018variational} and autoregressive prior \cite{minnen2018joint} models to estimate the entropy of the latent feature efficiently.
What's more, Ma \emph{et al.}\cite{ma2020end} introduce a special end-to-end framework for both lossy and lossless image compression, which adopts a trained wavelet-like transform to generate coefficients and a context-based entropy model to code those coefficients effectively.
Those works take the RD cost as the objective to end-to-end optimize the whole image coding frameworks with numerical methods.
Experiment results show that the latest end-to-end image coding method has achieved competitive performance with H.266, which fully verified the feasibility of end-to-end RD optimization with numerical methods.

Inspired by the successes in end-to-end image coding, some end-to-end learned video coding methods have been proposed\cite{chen2017deepcoder,wu2018video,chen2019learning,lu2019dvc,lu2020end,habibian2019video,djelouah2019neural,rippel2019learned,yilmaz2020end,yang2020learning,lin2020m,agustsson2020scale,liu2020neural,hu2021fvc}.
Different from image coding that removes the spatial redundant, video coding should also remove the temporal redundant. For this purpose, inter-picture prediction is a critical issue in video coding.
Accordingly, end-to-end video coding is more complex than image coding.
In 2018, Wu \emph{et al.}\cite{wu2018video} proposed a recurrent neural network (RNN) based video compression approach.
To code the B frame, it performs ME and then takes the obtained motion information to interpolate from two reference frames.
However, the ME in this work is not carried out by learned networks, but by classical search methods.
Therefore, this work is not end-to-end optimized.

Later on, a real end-to-end deep video compression framework (DVC)\cite{lu2019dvc,lu2020end} was proposed by Lu \emph{et al.} for the first time in 2019.
All modules in the framework are based on deep networks, including ME, MC, motion information compression, and residual compression modules.
The entire network is jointly offline optimized by numerical methods with a single loss function, i.e., the joint rate-distortion cost.
Based on this optimization idea, a series of end-to-end optimized video compression methods were proposed \cite{habibian2019video,djelouah2019neural,rippel2019learned,yilmaz2020end,yang2020learning,lin2020m,agustsson2020scale,liu2020neural,yang2020learningRLVC,hu2021fvc}.
For example, Yang \emph{et al.}\cite{yang2020learning} proposed a hierarchical end-to-end learned video compression (HLVC) for B-frame coding, which introduces hierarchical quality layers and a recurrent enhancement network to make full use of the temporal correlation.
Hu \emph{et al.}\cite{hu2021fvc} proposed an end-to-end optimized feature-space video coding framework (FVC). It performs all major operations (i.e. ME, MC, motion compression, and residual compression) in the feature space.

From those works, we can find that the mentioned methods have some same characteristics in optimization.
First, these offline designed models are based on continuous function entirely, so the optimization parameter space is continuous.
Second, the models are optimized only on the offline step, and they directly infer the coding results on the online step. They use the numerical method to optimize the continuous model offline.
Third, the optimization objective is the actual RD cost, so the optimization is end-to-end.

Based on all the above analyses in this section, we found that different kinds of coding frameworks use different optimization methods, and different optimization methods also have different advantages.
First, offline optimization can reduce the computational cost, and online optimization can achieve adaptive coding for each video.
Second, numerical optimization in continuous space is more efficient and refined in local parts, and search optimization in discrete space can find better coding parameters globally among multiple modes.
Third, it is more efficient to end-to-end optimize with the actual RD cost.
We can find that using only one optimization method can not achieve all the advantages.
Therefore, we would like to ask whether those optimization methods can be combined together to address this problem.

Based on this idea, a few coding methods have emerged\cite{lu2020content,hu2020improving}.
To make end-to-end learned coding adaptive for each video, Lu \emph{et al.} \cite{lu2020content} conduct online end-to-end numerical optimization on the offline end-to-end trained DVC model.
It optimizes the motion information (i.e. optical flow) with gradient-descent methods at the encoder side online without changing the decoder.
However, online gradient-descent optimization starting from only one group of offline trained model parameters may fail into the local minima.
Hu \emph{et al.}\cite{hu2020improving} add multi-resolution modes search optimization into offline end-to-end trained DVC to improve flow coding efficiency.
Based on the offline numerically optimized model, this work online searches for the mode with the lowest RD cost from a few resolution flow modes. The optimization in this method is end-to-end, i.e. with the actual RD cost.
However, this work performs numerical optimization only at the offline step, but doesn't conduct online to achieve more refined optimization.
Meanwhile, the number of modes is too few, making it hard to search for the globally optimal parameters.

\section{Theoretical Analysis of Optimization}
\label{sec_Theoretical Analysis}
\begin{figure}
  \centering
  \includegraphics[width=\columnwidth]{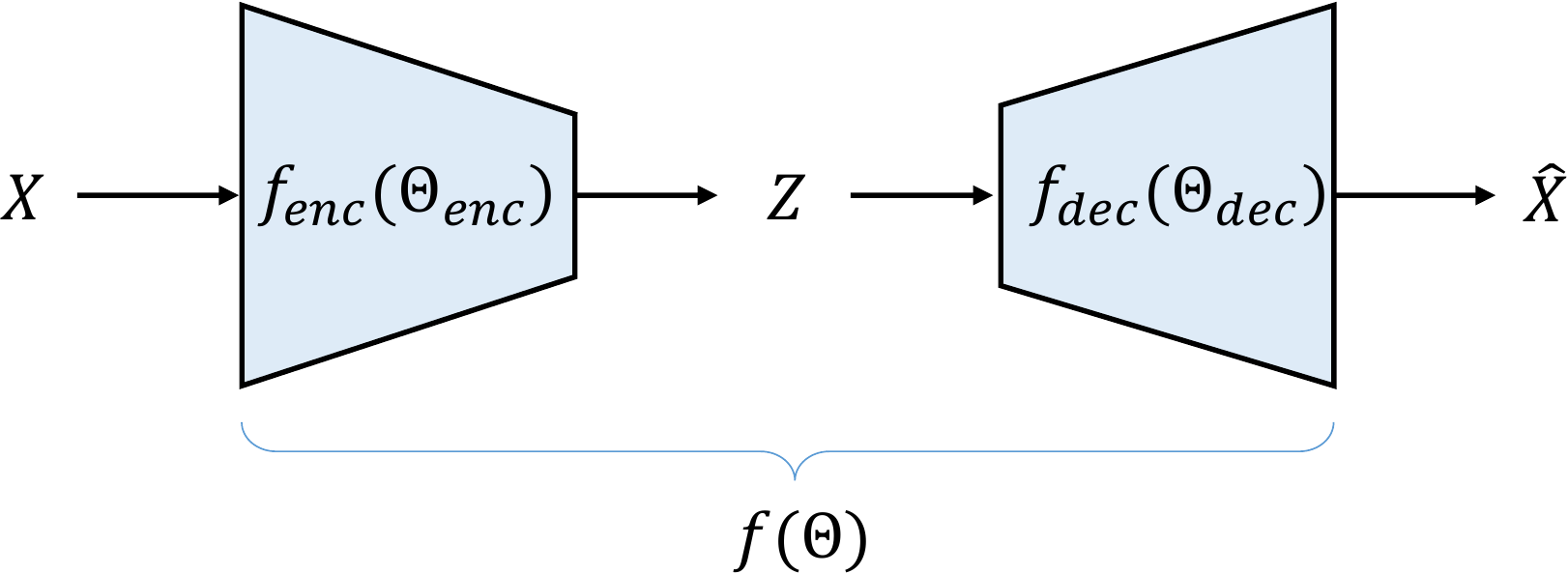}
\caption{Illustration of encoder-decoder in video coding.}
\label{fig Encoder-Decoder}
\end{figure}

\subsection{Optimization Problem Formulation}
\label{sec_Optimization Problem Formulation}
The theoretical foundations of lossy compression are rooted in Shannon's seminal work on rate-distortion theory \cite{Shannon1948A,Shannon1959Coding}.
Rate-distortion theory analyzes the fundamental tradeoff between the rate used for representing samples from a data source variable $X$, and the expected distortion incurred in decoding those samples from their compressed representations.
Formally, the relation between the input variable $X$ (original signal) and output variable $\hat{X}$  (compressed signal) is a (possibly stochastic) mapping defined by conditional distribution $P(\hat{X}|X)$.
The stochastic mapping $P(\hat{X}|X)$ is also called a test channel.
Based on those, Shannon's theory defined the classic optimization problem of rate-distortion.
That is, for a data source variable $X$, if the expected distortion is restricted to be bounded by $D_c$, then it optimizes to obtain the lowest rate
\begin{equation}
\mathop{\min}_{P(\hat{X}|X)}{I(X, \hat{X})} \text{ s.t. } \mathbb{E}[D(X, \hat{X})] \leq D_c
\label{equ_shannon_classical_RD_optimization}
\end{equation}
where $I$ denotes mutual information.
For the video coding problem, the test channel is implemented by an encoder-decoder pair $F$, i.e. $\hat{X} = F(X)$. The encoder-decoder pair $F$ is a deterministic mapping (a special case of stochastic mapping)
\begin{equation}
P(\hat{X}|X) = \left\{
\begin{array}{lr}
1,\  \rm{if}\ \hat{X} = F(X) &\\
0,\ \rm{otherwise}
\end{array}
\right.
\label{equ_P_transition probability}
\end{equation}
The input $X$ must be mapped to output $\hat{X}$ and cannot be mapped to other outputs, so the conditional probability $P(\hat{X}|X)$ is either $0$ or $1$.
Because the encoder-decoder space $F$ is huge, it is usually necessary to set the hyper-parameters of the encoder-decoder first, i.e. designing model structure $f$, to determine the optimized parameter space.
The hyper-parameters are not optimized, while the parameters $\Theta$ in the model structure need to be optimized in space $\Omega$. Thus, the $F$ can be converted to a parametrical expression $f(\Theta)$
\begin{equation}
F(X)\triangleq f(X|\Theta),\ \Theta \in \Omega
\label{equ_test channel parametrical expression}
\end{equation}
The model's parameters $\Theta$ consists of the encoder's parameters $\Theta_{enc}$ and the decoder's parameters $\Theta_{dec}$
\begin{equation}
{\Theta}= \Theta_{enc}\ \rm{and}\ \Theta_{dec}
\label{equ_optimization_parameters_consist}
\end{equation}
Accordingly, the optimization problem Eq.\eqref{equ_shannon_classical_RD_optimization} can be expressed as
\begin{equation}
\mathop{\min}_{\Theta \in \Omega}{I(X,\hat{X})} \text{ s.t. } \mathbb{E}[D(X, \hat{X})] \leq D_c
\label{equ_shannon_RD_optimization_for variable distribution}
\end{equation}
According to Shannon' information theory, the mutual information $I(X, \hat{X})$ is equal to the difference between the entropy $H(\hat{X})$ with the conditional entropy $H(\hat{X}|X)$.
Due to the definition of entropy and the deterministic mapping in Eq.\eqref{equ_P_transition probability}, the conditional entropy $H(\hat{X}|X)$ must be equal to $0$, i.e. $H(\hat{X}|X)=0$.
Thus, the the mutual information $I(X, \hat{X})$ is equal to the entropy $H(\hat{X})$
\begin{equation}
\begin{split}
I(X, \hat{X})&= H(\hat{X}) - H(\hat{X}|X) \\
&= H(\hat{X})
\end{split}
\label{equ_mutual information}
\end{equation}

In the practical coding, it uses the syntax to express the compressed signal. Specifically, the original signal $X$ is input into the encoder to generate the syntax $Z$, and then the syntax $Z$ is input into the decoder to obtain the compressed signal $\hat{X}$, as shown in Fig.\ref{fig Encoder-Decoder}.
\begin{equation}
Z = f_{enc}(X;\Theta_{enc});\hat{X} = f_{dec}(Z;\Theta_{dec})
\label{equ_mutual information}
\end{equation}
Specially, for the determined decoder $f_{dec}(\Theta_{dec})$, one syntax $Z$ products only one decided output $\hat{X}$, but one $\hat{X}$ may come from multiple syntax $Z$. Thus, $f_{dec}(\Theta_{dec})$ may be not one-to-one mapping (not injective function). Accordingly, $H(Z)$ is the upper bound of $H(\hat{X})$
\begin{equation}
H(\hat{X}) \leq H(Z)
\label{equ_mutual information}
\end{equation}
Since it is hard to measure $H(\hat{X})$, to minimize $H(\hat{X})$, we can change to minimize its upper bound $H(Z)$.
We denote the probability distribution of the variable $Z$ as $Q$, i.e. $Z \backsim Q$, and the entropy $H(Z)$ is
\begin{equation}
H(Z)= -\sum_{j}{Q(z_j})log Q(z_j)
\label{equ_mutual information}
\end{equation}
However, the true probability distribution $Q$ usually is unknown, so it uses an estimated probability distribution $B$ to approximate $Q$ to code $Z$.
\begin{equation}
\begin{split}
H(Q,B) &= -\sum_{j}{Q(z_j})log B(z_j) \\
& \geq H(Z)
\end{split}
\label{equ_mutual information}
\end{equation}
$H(Q,B)$ is called cross entropy of between two probability distributions $Q$ and $B$.
We can find that $H(Q,B)$ is the upper bound of $H(Z)$, so $H(Z)$ can be minimized by minimizing its upper bound $H(Q,B)$.
Accordingly, we convert optimizing mutual information into optimizing entropy.
On the other hand, the entropy represents the average bit rate.
Thus, in the coding problem, optimizing mutual information is equivalent to optimizing the bit rate.
We donate the calculation of bit rate as $R()$, and $R(Z)$ presents the bit rate of coding syntax $Z$, i.e. the cross entropy $H(Q,B)$.
Based on above formula derivation, the RD optimization problem Eq.\eqref{equ_shannon_RD_optimization_for variable distribution} can be expressed as
\begin{equation}
\mathop{\min}_{\Theta \in \Omega}{R(Z)} \text{ s.t. } \mathbb{E}[D(X, \hat{X})] \leq D_c
\label{equ_shannon_RD_optimization_Rate}
\end{equation}
For a data source variable $X$, if the expected distortion is restricted to be bounded by $D_c$, then the encoder-decoder model parameters $\Theta$ are optimized to obtain the lowest bit rate of coding syntax $Z$, where syntax $Z$ is the output of the encoder and used for decoding the compressed signal $\hat{X}$.

In the above derivation of Shannon's theory, the source is regarded as a random variable, which describes the source as a probability distribution. However, in practical video coding, the source is the samples of the distribution.
Due to this difference, to adapt to the practical samples' coding, some limited conditions are usually added to the theoretical formula.
Accordingly, the optimization problem will be changed.
According to the amount of coding samples, the practical optimization problem can be categorized into two groups.

The first is to optimize the encoder-decoder parameters $\Theta$ for lots of samples.
The set of samples can be considered to represent the distribution of $X$.
To minimize the loss on the set of samples, an average optimal encoder-decoder parameters $\Theta$ is optimized.
Specifically, a total of $K$ samples $x_1,\cdots,x_K$ generate $K$ groups of syntax $z_1,\cdots,z_K$ through the encoder, and then the syntax generate $K$ corresponding compressed samples $\hat{x}_1,\cdots,\hat{x}_K$, respectively
\begin{equation}
z_k=f_{enc}(x_k;\Theta_{enc});\hat{x}_k=f_{dec}(z_k;\Theta_{dec});k=1,\cdots,K
\label{equ_R_rate_function}
\end{equation}
For the set of samples, the rate part becomes the upper bound of bit rate for coding all samples
\begin{equation}
R(z_k) = -log B(z_k);k=1,\cdots,K
\label{equ_R_rate_function}
\end{equation}
The distortion part becomes
\begin{equation}
{\mathbb{E}[D(X, \hat{X})]|}_{X=x_1,\cdots,x_K;\hat{X}=\hat{x}_1,\cdots,\hat{x}_K} = \frac{1}{K}\sum_{k=1}^{K}{D(x_k, \hat{x}_k)}
\label{equ_D_distortion_function}
\end{equation}
Based on those equations, the RD optimization problem Eq.\eqref{equ_shannon_RD_optimization_Rate} can be expressed as
\begin{equation}
\mathop{\min}_{\Theta \in \Omega}{\sum_{k=1}^{K}{R(z_k)}} \text{ s.t. } \frac{1}{K}\sum_{k=1}^{K}{D(x_k, \hat{x}_k)} \leq D_c
\label{equ_shannon_RD_optimization}
\end{equation}
For lots of source samples, if the average distortion is restricted to be bounded by $D_c$, then it optimizes an average optimal encoder-decoder model parameter group $\Theta$ to obtain the lowest sum of bits of coding all samples' syntax.
The average optimal parameters can be directly deployed on the encoder/decoder side for encoding/decoding all the samples.

The second is to optimize for only one given sample.
Different from the average optimum for lots of samples, for each given sample, we can optimize a group of optimal encoder-decoder parameters $\Theta$.
However, as for the optimal decoder parameters $\Theta_{dec}$ for each sample, deploying parameters $\Theta_{dec}$ for all samples on the decoder side will take up lots of storage, while transmitting them to the decoder side will pay lots of bits.
Therefore, for the optimization of one given sample, it usually needs to fix the decoder parameters $\Theta_{dec}$.
It only optimizes the encoder parameters $\Theta_{enc}$ for each sample to match decoder parameters $\Theta_{dec}$ to achieve the best performance.
Specifically, a given sample $x$, the syntax $z$, compressed signal $\hat{x}$, rate $R(z)$, and distortion can be calculated by
\begin{equation}
\begin{split}
&z=f_{enc}(x;\Theta_{enc});\hat{x}=f_{dec}(z;\Theta_{dec})\\
&R(z) = -log B(z)\\ 
&{\mathbb{E}[D(X, \hat{X})]|}_{X=x,\hat{X}=\hat{x}} = D(x, \hat{x})
\end{split}
\label{equ_R_rate_function}
\end{equation}
Based on those equations, the RD optimization problem Eq.\eqref{equ_shannon_RD_optimization_Rate} can be expressed as
\begin{equation}
\mathop{\min}_{\Theta_{enc} \in \Omega_{enc}}{R(z)} \text{ s.t. } D(x, \hat{x}) \leq D_c
\label{equ_shannon_RD_optimization}
\end{equation}
where the $\Omega_{enc}$ is the optimization space of $\Theta_{enc}$.
For a given sample, if the distortion is restricted to be bounded by $D_c$, the decoder parameters $\Theta_{dec}$ is fixed and the encoder parameters $\Theta_{enc}$ are optimized to obtain the lowest bit rate of coding this sample's syntax.

To solve this kind of constrained optimization problem, the discrete version of Lagrange multipliers \cite{everett1963generalized} can be used to convert it into an unconstrained problem. The optimized goal is converted to minimizing the rate-distortion cost $J$ of the optimized parameters $\Theta$. Thus,
\begin{equation}
{\Theta}^{\ast} = {\mathop{\arg\min}_{\Theta \in \Omega}J},\ J = D + \lambda R
\label{equ_RD_optimization}
\end{equation}
where ${\Theta}^{\ast}$ donates the theoretically optimal parameters.

Although there are two coding optimization problems (i.e. for lots of samples or a single sample), the ultimate goal is to optimize the best parameters for each coding sequence. The optimization for lots of samples also serves for the optimization of a single sample.
Due to the complexity of coding problems, although the hyper-model is given, the parameter space for optimization is by far too large to be evaluated. Thus, the optimization is non-convex and difficult to be solved.
Over the years, a large number of studies have strived to solve the video coding optimization problem approximately.

\subsection{Optimization Problem Solution}
\label{sec_Optimization Problem Solution}
\begin{figure*}
  \centering
\subfigure[ ]
{
\includegraphics[width=0.3\linewidth]{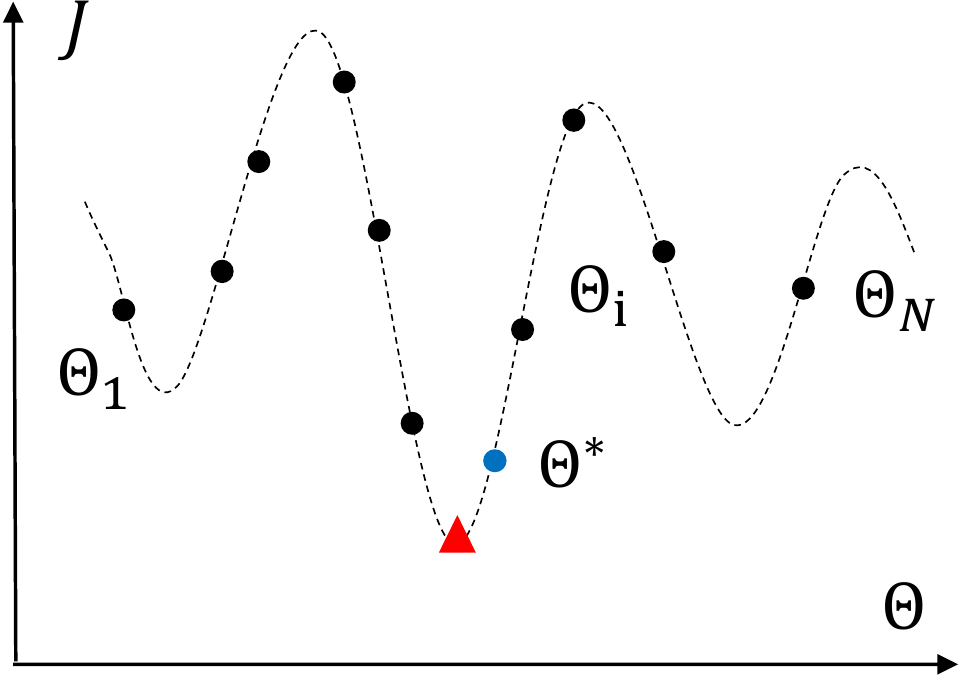}
}
\subfigure[ ]
{
\includegraphics[width=0.3\linewidth]{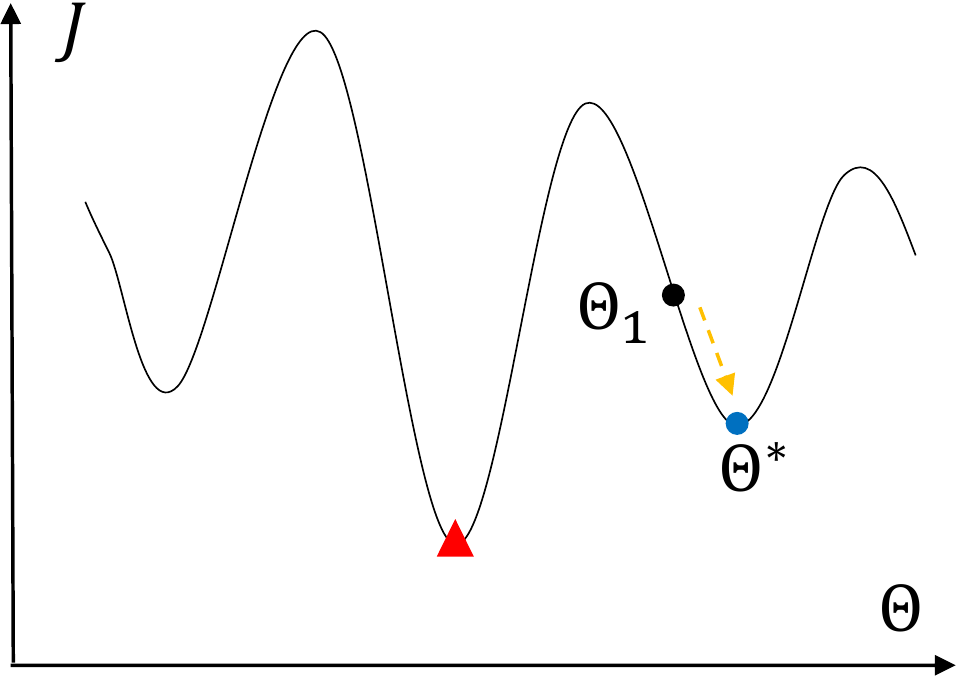}
}
\subfigure[ ]
{
\includegraphics[width=0.3\linewidth]{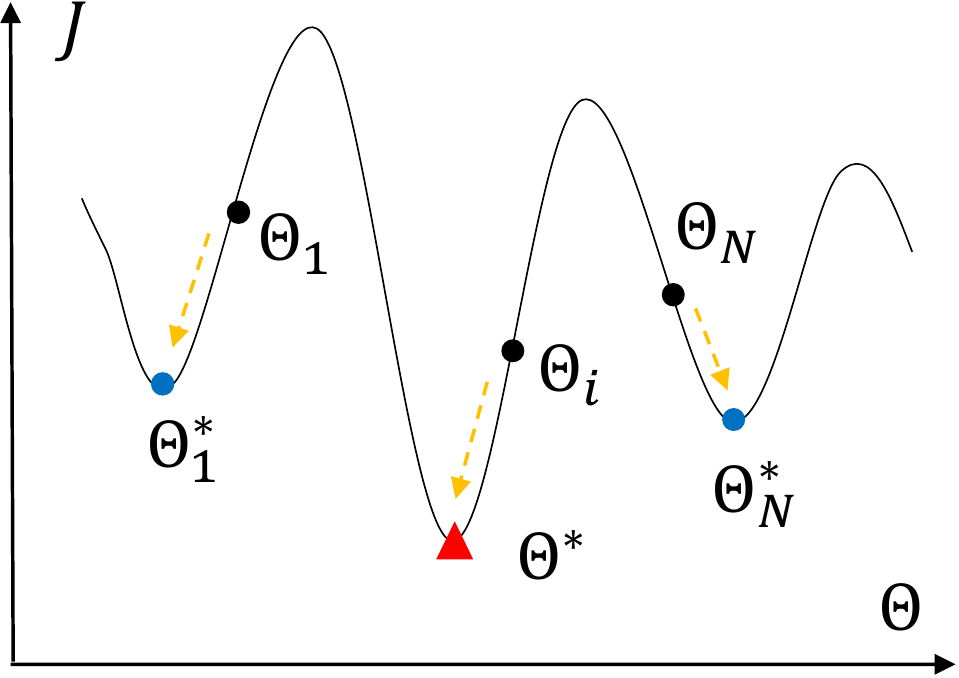}
}
\caption{Illustration of the optimization solution. (a) Search-based optimization in discrete space. (b) Numerical optimization in continuous space. (c) Proposed hybrid optimization.}
\label{fig Optimization_Analysis}
\end{figure*}

For given a video sequence $x$, the solution to the optimization problem of this samples is generally divided into two steps: offline determining the optimization space and optimizing the initial parameters through lots of samples, and online optimizing the parameters based on the offline model for the coded sample.
In the specific implementation, there are mainly two popular optimization solutions.

The one is still regarding the optimization of the test channel as a discrete optimization problem, and mainly using search-based optimization to solve it. The typical example is block-based hybrid coding.
To explain the optimization for a given data $x$ to be coded more clearly, we take the model's parameters $\Theta$ as the abscissa and the corresponding rate-distortion cost $J$ as the ordinate, and draw the simplified optimization diagram as shown in Fig.\ref{fig Optimization_Analysis} (a).

At the offline step, it designs a model structure to determine the whole discrete optimization space $\Gamma$ ($\Gamma \in \Omega$) firstly.
Discrete optimization space $\Gamma$ consists of dense parameter points by discretizing the continuous optimization space $\Omega$.
The discrete optimization space is so large that it is difficult to search from such a large number of parameter choices directly.
To address this problem, the discrete space usually needs to be further discretized (or quantized).
Specifically, the whole discrete optimization space $\Gamma$ is usually divided into $N$ non-overlapping subspaces $\Gamma_1, \cdots, \Gamma_N$
\begin{equation}
{\Gamma} = \Gamma_1 \cup \Gamma_2 \cup \cdots \cup \Gamma_N
\label{equ_shannon_RD_optimization}
\end{equation}
Each subspace $\Gamma_i$ is represented by a discrete optimal point $\Theta_i$ that searches through lots of training samples $x_1,\cdots,x_K$,
\begin{equation}
{\Theta_i} = \mathop{\arg\min}_{\Theta \in \Gamma_i}\sum_{k=1}^{K}{J(x_k;\Theta)}
\label{equ_shannon_RD_optimization}
\end{equation}
We take the union of all decoder parameters as the optimal decoder parameters $\Theta_{dec}^{\ast}$.
\begin{equation}
\Theta_{dec}^{\ast} = \Theta_{1\_dec} \cup \Theta_{2\_dec} \cup \cdots \cup \Theta_{N\_dec}
\label{equ_shannon_RD_optimization}
\end{equation}
Then the whole discrete optimization space is expressed by a set $\Psi$ with $N$ discrete starting points $\Theta_{1\_enc}, \cdots, \Theta_{N\_enc}$
\begin{equation}
\Psi = \{{\Theta_{i\_enc}}\ |i=1,2,\cdots,N\}
\label{equ_shannon_RD_optimization}
\end{equation}
At the online step, it searches the best encoder parameters $\Theta_{enc}^\ast$ from the set $\Psi$ for the coded sample $x$
\begin{equation}
{\Theta}_{enc}^{\ast} = \mathop{\arg\min}_{\Theta_{enc} \in \Psi}{J(x;\Theta)}
\label{equ_optimization_online_traditional}
\end{equation}

In the specific implementation of block-based hybrid coding, the two optimization steps are realized by offline designing multiple coding modes and setting mode parameters, and online searching.
The discrete optimized space $\Gamma$ is the finite range of all the model's parameters, and is divided into $N$ subspaces by set $N$ modes. There are still many options for parameters within subspace, and then it chooses the parameter $\Theta_i$ as the model's parameters in $i$-th subspace (e.g. for a quarter MV x-dimension value $0.25$, its discrete subspace contains any discrete value in the interval $[0.125, 0.375]$, and it finally chooses the value $0.25$ to represent this subspace).
The search-based optimization searches from multiple starting points globally so that it's more likely to optimize the best model parameters. However, the computational efficiency of search optimization is low. In addition, limited by the discreteness of the model's parameters and the search complexity, the search point is not dense enough to perform finer optimization, so it may not achieve better optimization results in local parts.

The other one is formulating the optimization of model parameters as a continuous optimization problem, and solving it with the numerical optimization method. End-to-end learned coding is a typical example. The simplified optimization diagram is shown in Fig.\ref{fig Optimization_Analysis} (b). At the offline step, it designs continuous functions to modeling the optimization space $\Omega$ and then establishes the continuous relationship between the optimized parameters $\Theta$ and RD cost $J$. Through minimizing the cost of lots of samples with the numerical optimization, a group of average optimal model's parameters is obtained
\begin{equation}
{\Theta_1} = \mathop{\arg\min}_{\Theta \in \Omega}\sum_{k=1}^{K}{J(x_k;\Theta)}
\label{equ_shannon_RD_optimization}
\end{equation}
where it only optimizes only one group of average optimal model parameters $\Theta_1$.
We take the decoder parameters $\Theta_{1\_dec}$ as the optimal decoder parameters $\Theta_{dec}^{\ast}$.
At the online step, for the coded sample $x$, it directly applies the initial parameters $\Theta_{1\_enc}$ or further optimizes a group of better parameters around $\Theta_{1\_enc}$ with the numerical optimization
\begin{equation}
{\Theta}_{enc}^{\ast} = \mathop{\arg\min}_{\Theta_{enc} \in \Omega_{\Theta_{1\_enc}}}{J(x;\Theta)}
\label{equ_optimization_online_traditional}
\end{equation}
where $\Omega_{\Theta_{1\_enc}}$ represents the convex space where $\Theta_{1\_enc}$ is.
In the specific implementation of end-to-end learned coding, the two optimization steps are achieved by offline designing a network model and training it with the gradient descent algorithm, and online direct inference or further fine-tuning around the offline trained model. The continuous optimized space $\Omega$ is the infinite range of all network parameters, $\Theta$ is the all network parameters, and $\Theta_1$ is the offline optimized network parameters.
The numerical optimization in the continuous space can not only solve the optimization problem more efficiently but also perform finer optimization to make it more likely to achieve the optimum in the local part. However, most of the numerical optimization are greedy algorithms, such as the gradient descent algorithm. Those schemes rely heavily on the starting point since they conduct optimization around it. Only one starting point will fall into a local optimum.

Based on the two optimization solutions, we propose a hybrid of discrete and continuous optimization to solve this problem efficiently. We provide multiple discrete start points in global continuous space and adopt numerical optimization in the local part around each starting point. Then we use search-based optimization among the multiple discrete local optimums to obtain the global optimum. The hybrid optimization diagram is shown in Fig.\ref{fig Optimization_Analysis} (c).
At the offline step, we first formulate the problem as continuous optimization and adopt numerical optimization on the continuous space $\Omega$ of the designed model.
We divide the whole continuous optimization space into $N$ non-overlapping continuous subspaces $\Omega_1, \cdots, \Omega_N$
\begin{equation}
{\Omega} = \Omega_1 \cup \Omega_2 \cup \cdots \cup \Omega_N
\label{equ_shannon_RD_optimization}
\end{equation}
We optimize an optimal point $\Theta_i$ for each subspace $\Omega_i$ through lots of training samples
\begin{equation}
{\Theta_i} = \mathop{\arg\min}_{\Theta \in \Omega_i}\sum_{k=1}^{K}{J(x_k;\Theta)}
\label{equ_shannon_RD_optimization}
\end{equation}
We take the union of all decoder parameters as the optimal decoder parameters $\Theta_{dec}^{\ast}$
\begin{equation}
\Theta_{dec}^{\ast} = \Theta_{1\_dec} \cup \Theta_{2\_dec} \cup \cdots \cup \Theta_{N\_dec}
\label{equ_shannon_RD_optimization}
\end{equation}
At the online step, for the coded sample $x$, we adopt numerical optimization to find each group of local optimal model's parameters $\Theta_i^{\ast}$ around each starting point $\Theta_i$
\begin{equation}
\Theta_{i\_enc}^{\ast} = \mathop{\arg\min}_{\Theta_{enc} \in \Omega_{\Theta_{i\_enc}}}{J(x;\Theta)}
\label{equ_optimization_online_traditional}
\end{equation}
where $\Omega_{\Theta_i}$ represents the convex space where $\Theta_i$ is. Then all $N$ discrete local optimums $\Theta_1^{\ast}, \cdots, \Theta_N^{\ast}$ form a set $\Psi$
\begin{equation}
\Psi = \{{\Theta_{i\_enc}^{\ast}}\ |i=1,2,\cdots,N\}
\label{equ_shannon_RD_optimization}
\end{equation}
and search-based optimization is used to search the global optimal one among all local optimums
\begin{equation}
\Theta_{enc}^{\ast} = \mathop{\arg\min}_{\Theta_{enc} \in \Psi}{J(x;\Theta)}
\label{equ_optimization_online_traditional}
\end{equation}
Optimization in the continuous space can optimize the local optimal parameters more efficiently. Meanwhile, the optimization in the discrete space can search for the best one from multiple discrete points globally, which can avoid the final solution falling into the local optimal as much as possible. Thus, proposed hybrid optimization is a better way to solve this problem.

\section{Proposed Method}
\label{sec_Proposed Method}
\begin{figure*}
  \centering
\subfigure[ ]
{
\includegraphics[width=0.45\linewidth]{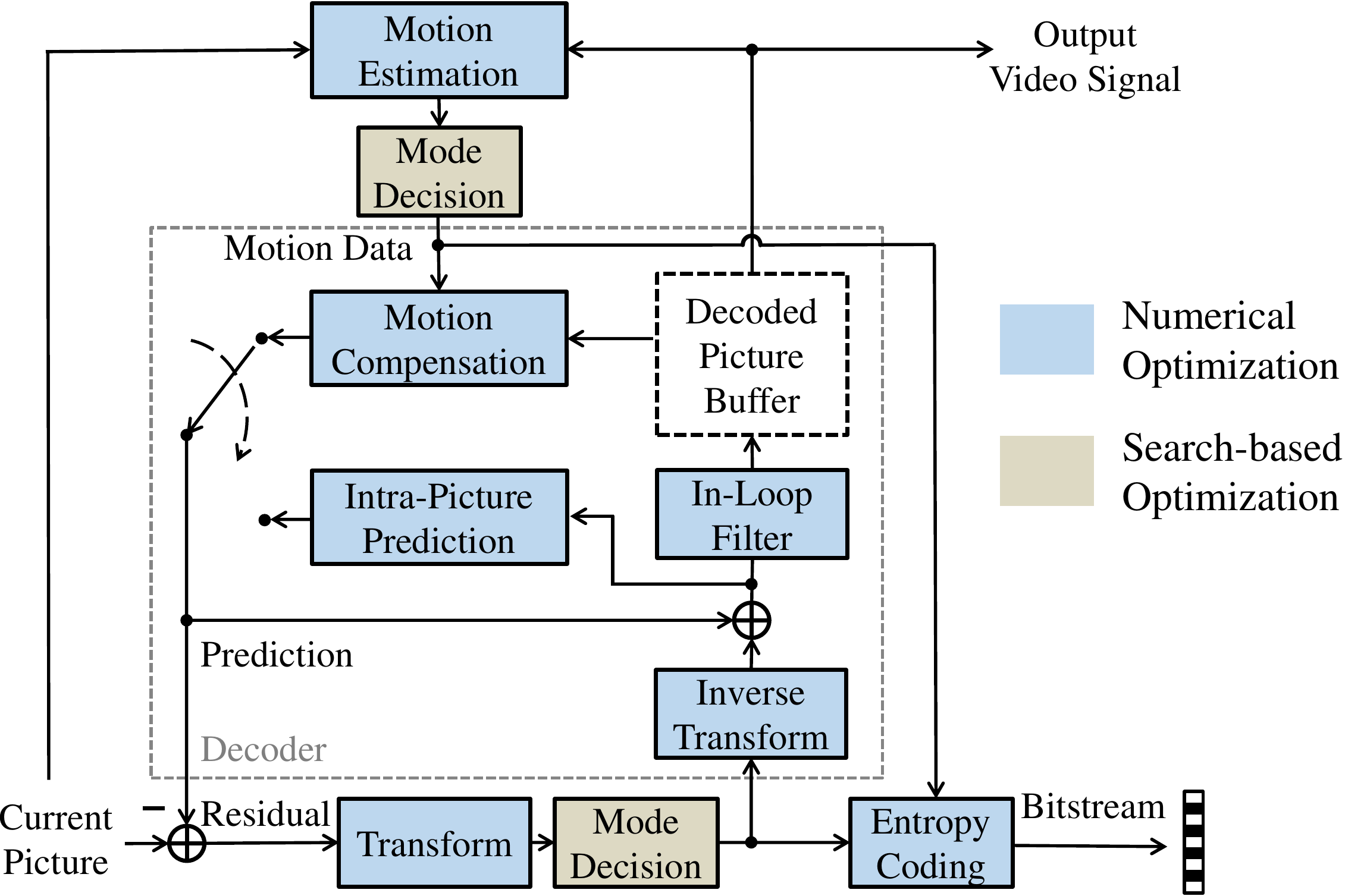}
}
\subfigure[ ]
{
\includegraphics[width=0.51\linewidth]{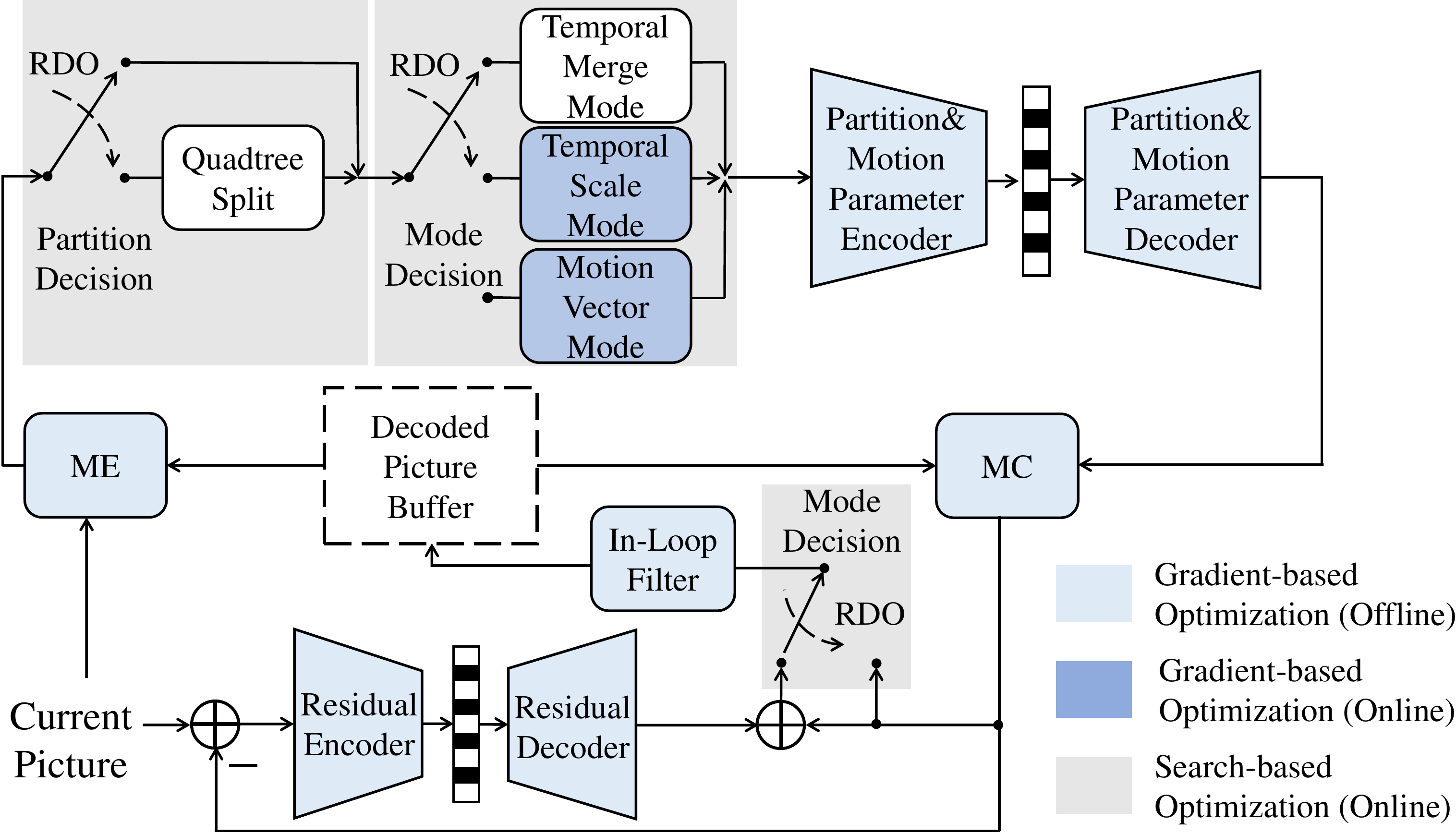}
}
\caption{(a) Overview of our hybrid-optimization video coding framework. (b) Implementation of our hybrid-optimization video coding framework.}
\label{fig Video_Compression_Framework}
\end{figure*}

Guided by the proposed hybrid-optimization theory, we propose a hybrid-optimization video coding method, as show in Fig.\ref{fig Video_Compression_Framework} (a).

\subsection{Offline Hybrid Optimization}
\label{sec_method_offline}
As above analysis, firstly, we offline set some hyper-parameters to determine an encoder-decoder model structure. 
Specifically, the hyper-parameters mainly consist of two parts.
The first is to determine the structure of the continuous part.
We adopt the hybrid framework of motion-handling and picture-coding techniques, which has proven very effective.
It consists of some modules, including motion estimation (ME), motion compensation (MC), transform/inverse transform, quantization/inverse quantization, in-loop filter, and entropy coding.
Then we set the hyper-parameters for each module to determine its internal structure.
We build those modules with continuously optimized tools (e.g. deep neural networks) so that they can be optimized with numerical methods.
The second is to determine the structure of the discrete part.
We design various discrete modes, which can divide the whole continuous optimization space into multiply sub-spaces. Due to the hybrid framework of motion-handling and residual picture-coding, the designed modes focus on the motion and transform coding accordingly. Through different motion models and residual coding methods, different content in the video can be handled efficiently.

Next, we offline optimize the model parameters through lots of training samples.
Specifically, the optimization can be divided into two steps.
First, we optimize the parameters of the whole continuous framework. We use numerical methods to optimize all the modules in an end-to-end manner.
Second, based on the optimized parameters of the continuous framework, we optimize the initial parameters of each discrete mode.
We convert the parameter format of the continuously optimized model to the format of each discrete mode, and use the converted parameters to initialize each mode.

Our proposed method is consistent with the proposed hybrid-optimization theory.
The continuous optimization space $\Omega$ is the range of all module and mode parameters.
The modes divide the continuous optimization space $\Omega$ into multiply sub-spaces $\Omega_1, \cdots, \Omega_N$.
Through offline optimization for all modes, we obtain multiply groups of initial mode parameters as the discrete starting points $\Theta_1, \cdots, \Theta_N$.

\subsection{Online Hybrid Optimization}
\label{sec_method_online}
Based on our offline optimized framework, we conduct online hybrid optimization for each frame to be coded.
Guided by the proposed hybrid optimization theory, the online optimization process is divided into two steps: continuous optimization in local parts and discrete optimization in global space.
Firstly, in the local area of each mode, we perform continuous numerical optimization around the initial parameters (i.e. $\Theta_i$) to obtain the locally optimal parameters (i.e. $\Theta_i^{\ast}$).
We conduct this optimization for all modes, including the motion-handling and transform coding mode, and then we can obtain the locally optimal parameters of all discrete modes (i.e. $\Psi$).
Secondly, we perform discrete search optimization among all locally optimal modes to select the globally best mode (i.e. $\Theta^{\ast}$).
Note that the online optimization is also end-to-end.
As a result, through the hybrid optimization for the current frame, different content in the video can be coded adaptively and efficiently.

\section{Implementation}
\label{sec_Implementation}
We design a deep network-based video compression framework to implement our proposed hybrid-optimization method.
The implementation is shown in Fig.\ref{fig Video_Compression_Framework} (b).
All modules of our framework are based on deep networks.

At the encoder, we use the optical flow network to conduct motion estimation between the current frame and reference frames and take the optical flow as the initial motion information. Then we design various partition modes and multi motion modes to describe the optical flow efficiently so that the expensive pixel-level flow is changed into low-cost partition and motion parameters.
We employ the motion compensation network to obtained the predicted frame from the reference frames and motion parameters.
The transform module is used to converts the residual into coefficients that are easier to code.
The transformed coefficients are determined whether to code or not by residual skip mode.
After that, we refine the reconstructed frame by the learned in-loop filter and output the final reconstruction.
The entropy encoding network generates the probability distribution of quantized motion parameters and transformed coefficients, and then codes them into the bitstream by an arithmetic encoder (AE).
At the decoder, with the help of an arithmetic decoder (AD), the entropy decoding network decompresses the motion parameters and transformed coefficients from the bitstream. Next, we sequentially use motion compensation, inverse transform, and in-loop filter to obtain the decoded reconstruction.

In this section, we will introduce the used coding configuration and our proposed coding modules in detail.

\subsection{Coding Configuration}
\label{sec_Coding Configuration}
In this paper, we adopt the random-access (RA) coding configuration, and follow the classical hierarchical B layers coding structure.
We refer these hierarchical B layers in the GOP as layer 0, 1, 2, 3, and so on, respectively in decoding order.
The last B frame in GOP (layer 0) only can perform uni-directional inter-prediction. However, other B frames can perform bi-directional inter-prediction by referring to the two reconstructed frames of the previous layer.
Considering this difference, in this paper, we only code the bi-directional B frames in the high layers by our designed deep networks, and the I frame and the B frames of layer 0 are coded by HEVC.
This configuration can verify the effect of our proposed method because the layer 0 frames only account for a small part of the video, that is, one B frame in one GOP and one I frame in one second.
Most of the frames are the B frames in the high layers.

\subsection{Motion Estimation}
\label{sec_Motion Estimation}
Motion estimation in block-based hybrid video coding uses search-based optimization to online search for the best motion field at the encoder side, which usually takes lots of time.
However, learning-based optical flow can infer a motion field directly with the offline trained model.
In our framework, we adopt the optical flow network PWC-Net\cite{sun2018pwc} to estimate an initial motion field rapidly. The motion field includes the optical flow between the current frame and reference frames and the optical flow between reference frames. These optical flows will be used for later optimization.

\subsection{Block Partition}
\label{sec_Block Partition}
The correlation between the adjacent pixels in a frame is usually strong, while the correlation between the far away pixels may be weak. In order to reduce this redundancy, block partition is widely used in video coding. The main purpose of block partition is to group samples that can be processed efficiently and are coded together.
In our framework, we adopt the variable block partition optimization that provides multiple block size modes to enable high efficiency and adaptation to different regions.
Specifically, we use quadtree-structured partitioning and signaling like HEVC.
Only square block partitioning is specified, and a block can be recursively split into four quadrants.
Each quadrant is assigned a flag that indicates whether to split.
The partitioned block includes four sizes: $8\times8$, $16\times16$, $32\times32$, $64\times64$, and the partitioned result is described efficiently by the split flag.
Finally, through the flexible partition and the parameter sharing within a block, we avoid the expensive overhead of pixel-level flow and achieve high-efficiency compression.

\subsection{Motion Mode}
\label{sec_Motion Mode}
The efficient representation of motion is the key to improving video coding performance.
Block-based hybrid video coding usually describes motion at a high level, which means the motion model is designed according to the motion type, such as the block translation model and the affine model for rotation and zooming. However, the motion field is usually not as fine as at the pixel level.
On the other hand, optical flow in the learned end-to-end video coding is a more straightforward way to describe motion at the pixel level, but it ignores high-level motion characteristics.
Different motion models are beneficial to different motion situations. Therefore, in our framework, we introduce a hybrid of different motion models and propose multi-mode optimization to deal with different motion efficiently. We design three motion modes: motion vector (MV) mode, temporal merge (TMerge) mode, and temporal scale (TScale) mode. We signal a mode index for each block to decide which mode is used in this block. Note that we will convert the coded block-level motion parameters into the pixel-level optical flow field for later motion compensation operation.

Specifically, only a MV can describe the translation of a block in block-based hybrid video coding. Similarly, we also adopt motion vector mode in our framework to handle the translational motion. Based on the bi-directional reference frames at time $t-k$ and $t+k$, the bi-directional optical flow of all pixels in a block at time $t$ are set as $f_{t}$
\begin{equation}
f_{t}=\{mv_{t \rightarrow t-k}, mv_{t \rightarrow t+k}\}
\label{equ_MV}
\end{equation}
where $mv_{t \rightarrow t-k}$ and $mv_{t \rightarrow t+k}$ denote two motion vectors for bi-direction, and they are quantized with half-pixel precision to code.
In addition, like learned end-to-end video coding, we also adopt pixel-level optical flow to deal with finer motion. However, it's not efficient to transmit the optical flow directly because the motion information of bi-directional reference frames provides lots of prior information for the current frame prediction. To make full use of the prior motion information, we propose temporal merge mode that can utilize the pixel-level flow without transmitting any motion information. Since the time interval between the current frame and reference frames is usually very short, the motion in most regions can be considered uniform. Based on the uniform motion assumption and the prior optical flow between reference frames $v_{t-k \rightarrow t+k}$, $v_{t+k \rightarrow t-k}$, we can obtain the bi-directional pixel-level flow field in a block by
\begin{equation}
f_{t}=\{v_{t+k \rightarrow t-k} \times 0.5, v_{t-k \rightarrow t+k} \times 0.5\}
\label{equ_TMerge}
\end{equation}
Due to the diversity of motion, the motion in some regions is not uniform, but it may obey some rules, such as accelerated motion. Inspired by both coding frameworks, we propose a hybrid mode of high-level motion parameters and pixel-level flow to address the problem. This hybrid mode, called temporal scale mode, combines the high efficiency of the high-level parameter-based motion model and the fine precision of the flow-based motion model.
In our framework, we use a group of scaling parameters to describe the temporally characteristic of motion and then combine the parameters with the prior flow to generate the bi-directional pixel-level flow field:
\begin{equation}
f_{t}=\{v_{t+k \rightarrow t-k} \times s_{0}, v_{t-k \rightarrow t+k} \times s_{1}\}
\label{equ_TScale}
\end{equation}
where $s_{0}$ and $s_{1}$ denote two two-dimensional scaling vectors for bi-direction, and they are quantized with one-tenth precision. Finally, the motion of a block is signaled by four scaling values.

Our proposed multiple modes can handle most motions effectively. According to the uniformity of motion in spatio-temporal domain, the motion can be classified into four categories: spatially uniform motion, spatially non-uniform motion, temporally uniform motion, and temporally non-uniform motion. Since the proposed block partition mode can adaptively divide the pixels with the same motion into one block and the pixels with different motion into different blocks, it can deal with the spatially uniform and non-uniform motion to a certain extent. The proposed temporal merge mode is specially designed for temporally uniform motion. As for temporally non-uniform motion, if the motion is regular and the prior flow is estimated accurately, the proposed temporal scale mode can handle it well. Otherwise, the motion vector mode by transmitting block-based MV directly will provide a good approximation for the motion.

\subsection{Motion Compensation}
\label{sec_Motion Compensation}
\begin{figure}
  \centering
  \includegraphics[width=\columnwidth]{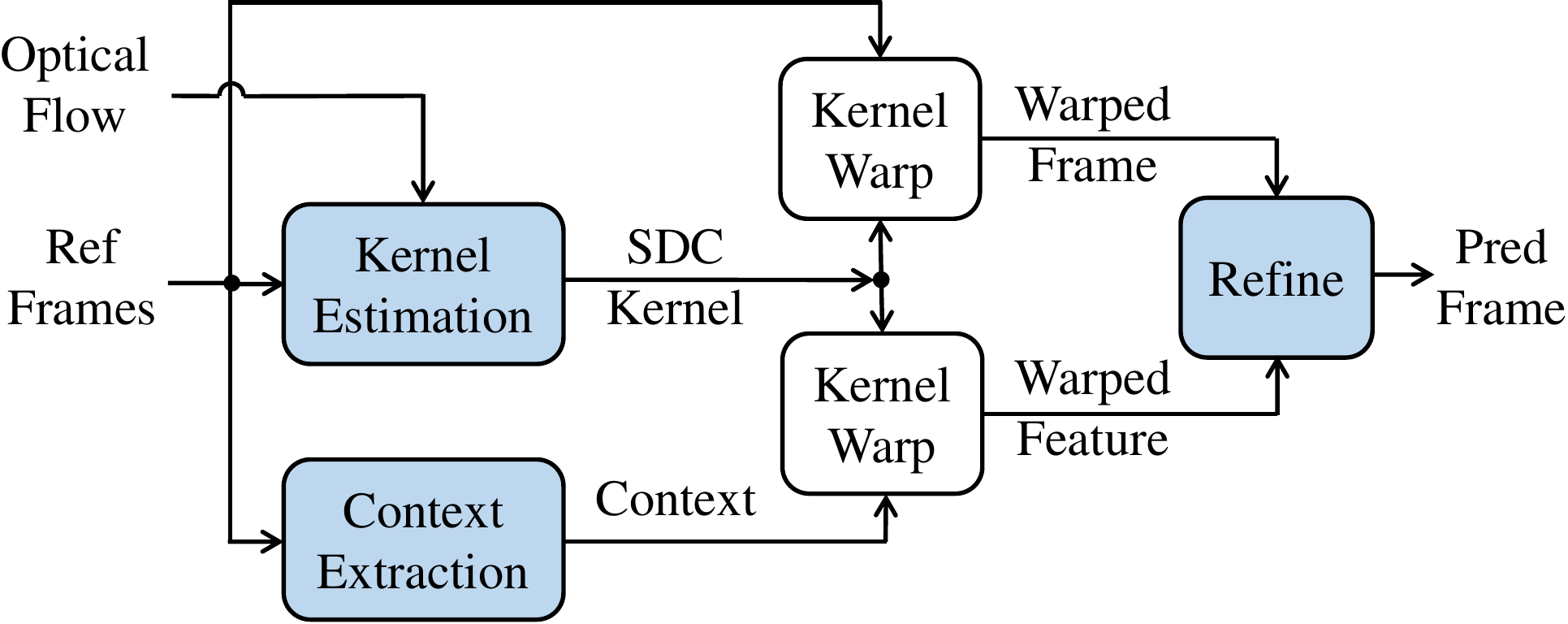}
  \caption{Motion compensation network.}
  \label{fig Motion_Compensation_Network}
\end{figure}

Motion compensation is to retrieve the predicted frame from the reference frames through the motion field.
Previous video coding methods usually use the fixed fractional-pixel interpolation filter in MC, such as DCTIF in classical HEVC and warping operation in learned DVC.
In this paper, we adopt an adaptive kernel-based MC network.
The kernel-based MC that synthesis a pixel by convolving input patches with a learned kernel \cite{niklaus2017videosep} can make full use of spatial information and conduct adaptive prediction for each pixel.
We use spatially-displaced convolution (SDC)\cite{reda2018sdc} which applies the learned kernel at the location of the motion field to combine the given motion field and learned kernel effectively.
Besides, we utilize the contextual information in the feature domain to improve MC prediction quality further\cite{niklaus2018context,bao2019memc}.
Our MC network is shown in Fig.\ref{fig Motion_Compensation_Network}.
We learn a kernel for each pixel and synthesize a pixel by applying the kernel on the reference frame at the displaced location of the given motion field.
We extract the context from the pre-trained ResNet18 \emph{conv1} layer\cite{he2016deep} and also apply the learned SDC kernel to warp the feature. A refine network is adopted to fuse them to generate the final MC prediction frame.

\subsection{Transform and Quantization}
\label{sec_Transform and Quantization}

\begin{figure}
  \centering
  \includegraphics[width=\columnwidth]{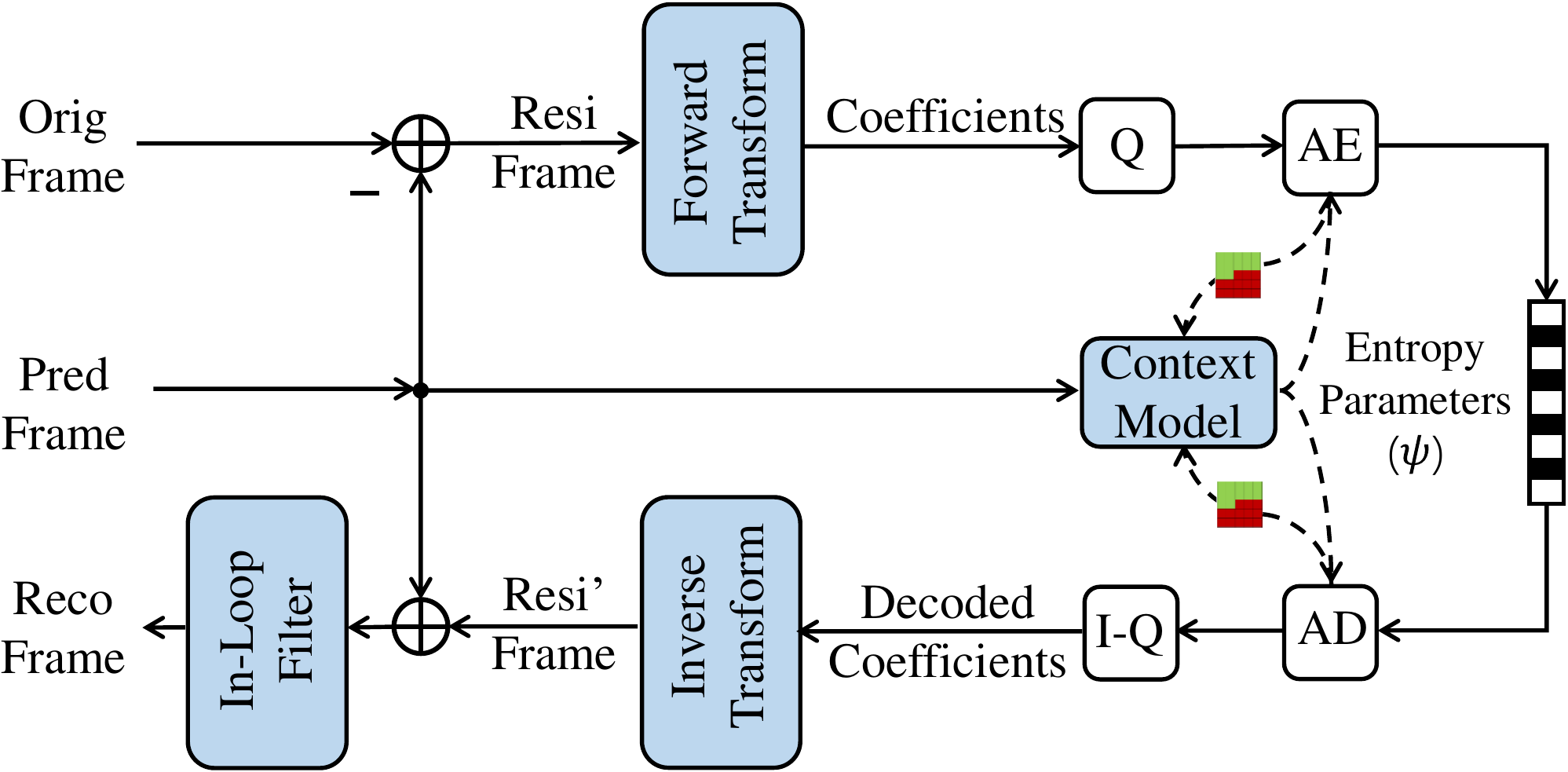}
  \caption{Residual coding network.}
  \label{fig Residual_Coding_Network}
\end{figure}

Transform aims to convert the residual into coefficients that are easier to code, and quantization is used to reduce the precision of coefficients to decrease the amount of data.
Correspondingly, inverse quantization restores the original coefficient range without regaining the precision, and inverse transform is applied to reconstruct the residual samples from the restored coefficients.
In our framework, we adopt wavelet transform/inverse-transform\cite{christopoulos2000jpeg2000} at a frame level and uniform scalar quantization/inverse-quantization.
Our residual coding network is shown in Fig.\ref{fig Residual_Coding_Network}.
Wavelet transform results in a plurality of subbands of different resolutions.
Similar to JPEG2000\cite{christopoulos2000jpeg2000}, we perform 4 times of pyramid decomposition in wavelet transform, i.e. the residual frame is forward transformed into four subbands $\{LL_{1}, HL_{1}, LH_{1}, HH_{1}\}$ at first and then we conduct the $k$-th forward transform from $LL_{k-1}$ into $\{LL_{k}, HL_{k}, LH_{k}, HH_{k}\}$ recursively. Here, $LL$, $HL$, $LH$, $HH$ are all two dimensional; $L$ stands for low-frequency and $H$ stands for high-frequency.
Finally, we got 13 subbands with coefficients.
At the encoder side, all the forward transformed coefficients $y$ are scalar quantized to obtain the discrete indexes $q$ to be coded:
\begin{equation}
q=[y / Q_{step}]
\label{equ_Q}
\end{equation}
where $[ ]$ stands for rounding operation and the unique parameter $Q_{step}$ is used to control the rate and distortion.
At the decoder side, the reconstructed coefficients $\hat{y}$ are obtained by uniform scalar inverse-quantization:
\begin{equation}
\hat{y}= q \times Q_{step}
\label{equ_I-Q}
\end{equation}
The inverse transform is strictly the reverse of the forward transform, which is realized by reversing the order of the operations and swapping the additions and subtractions.

\begin{figure}
  \centering
  \includegraphics[width=\columnwidth]{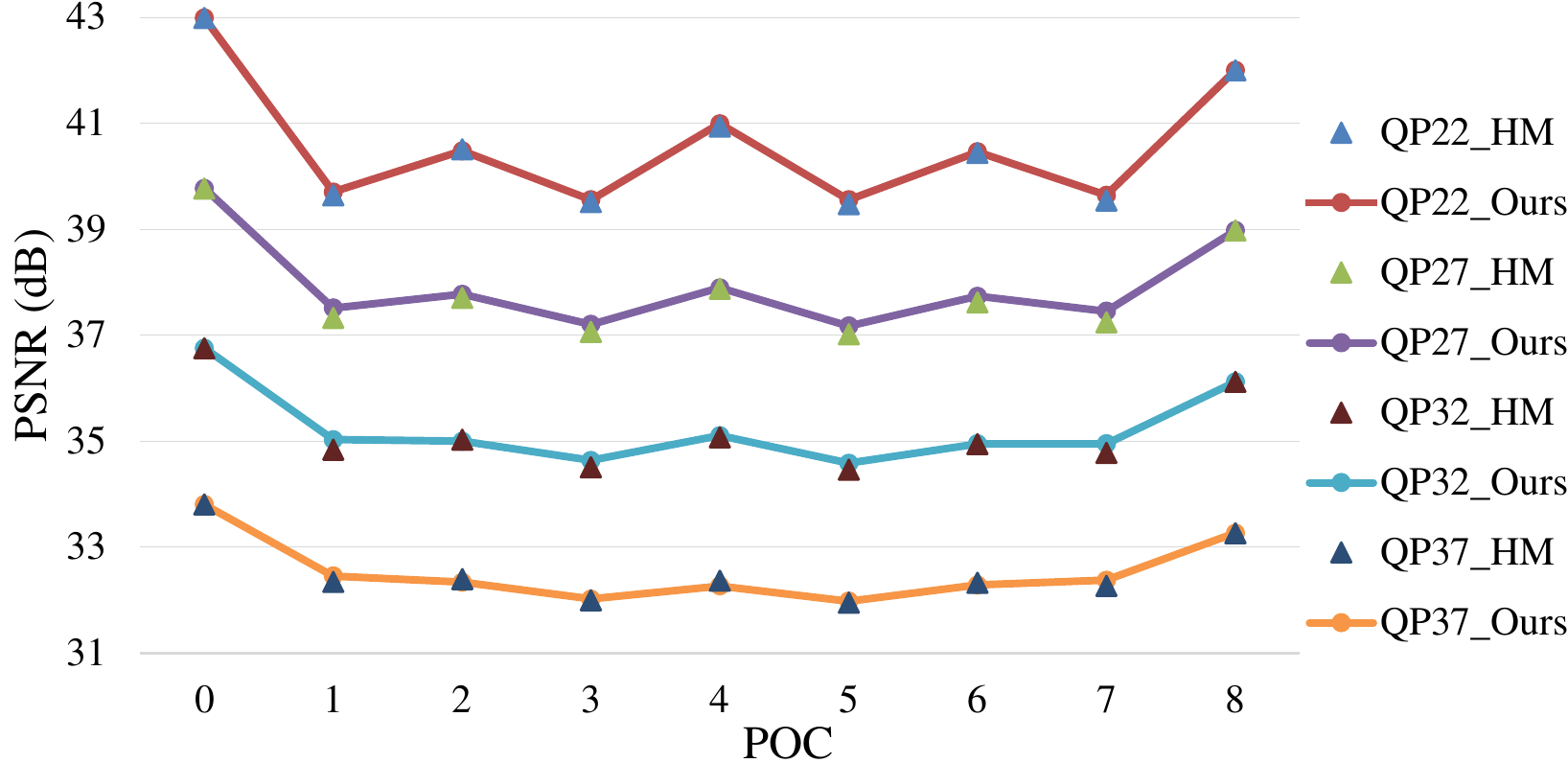}
  \caption{Quality comparison of each reconstructed frame in the Group Of Picture (GOP) between HM and our proposal. The PSNR is the average result of all GOPs of all sequences.}
  \label{fig Hierarchical_Structure}
\end{figure}

Our framework adopts wavelet transform mainly because it has a clearer correspondence between transformed coefficients and pixels. The bits of each partitioned block can be estimated approximatively from coefficients. Thus, it's very suitable for block-based rate-distortion optimization among multiple modes in our framework.
Moreover, we can achieve virtually any reachable bitrate and quality by simply adjusting the quantization step of wavelet transform coefficients. As shown in Fig.\ref{fig Hierarchical_Structure}, the quality of the reconstructed frames can be controlled precisely to make it almost the same as HM. Therefore, we not only can reach hierarchical quality easily but also can solve the problem of error propagation more effectively compared to other learned video coding methods\cite{lu2020content}.
In addition to lossy compression, since the wavelet transform doesn't lose any information, the coding methods based on wavelet transform can also support lossless compression by transforming in the integer domain\cite{christopoulos2000jpeg2000}.

\subsection{Residual Mode}
\label{sec_Residual Mode}
The coding performance of different videos may benefit from different residual coding methods.
For example, when the prediction is very close to the original signal, coding the sparse residual will decrease the coding efficiency. To address this problem, we propose residual skip (ResiSkip) mode that can skip the coding of transformed coefficients in some regions. We divide the residual frame into residual-skip units regularly, and we signal a flag for each unit to decide whether skip the residual coefficients of this unit. The residual-skip unit is set to $128\times128$ to avoid too much overhead.
Specifically, we transform the residual frame into coefficients firstly. If the unit chooses residual skip mode, we set the corresponding coefficients of this unit to 0 and skip the coding of those coefficients. Otherwise, we code the coefficients. The residual skip mode can not only improve the coding efficiency in some cases, but also avoid the computation of coding coefficients.

\subsection{In-Loop Filter}
\label{sec_Loop-Filter}
After MC and residual compression, we can reconstruct the current frame from the prediction and residual frame. However, the reconstruction contains compression error, especially at low bit rates. It not only reduces the reconstruction quality, but also affects the inter-prediction of the next coded frame due to the reference frame mechanism. Therefore, we propose an in-loop filter network to compensate for the compression error of the reconstructed frame.
It is operated within the inter-picture prediction loop.
We concatenate the reconstructed frame, predicted frame, and residual frame as the input to make full use of the spatio-temporal correlation, then feed them into a CNN with eight residual blocks\cite{he2016deep} to obtain the refined reconstructed frame.

\subsection{Entropy Coding}
\label{sec_Entropy Coding}
In this paper, we adopt the context-based entropy coding method similar to \cite{ma2020end} and the factorized entropy coding method.
The factorized entropy coding uses the fixed statistical probability model.
The context-based entropy coding predicts a probability distribution for each coefficient, where the probability distribution is constructed by a set of entropy parameters $\psi$ of CNN output, as shown in Fig.\ref{fig Residual_Coding_Network}. Arithmetic coding is used to code each coefficient according to its probability distribution.
The statistical probability model is used to code the parameters with weak correlation to reduce computational complexity, while the context model is used to code the parameters with strong correlation to improve coding performance.
We use the statistical probability model to code split flag, motion mode index, residual skip mode flag, and the scaling parameters of temporal scale mode. The MVs of motion vector mode are coded directly by the CNN-based context model.
As for the transformed coefficients, we adopt the CNN-based context model to exploit the correlation among prediction signal, wavelet coefficients within and across subbands.
When coding one wavelet coefficient, CNN generates its conditional probability distribution by inputting three context information: the already coded coefficients in the current subband, other already coded subbands, and the inter-prediction signal.
Note that we do not perform real arithmetic encoding during optimization but instead estimate the bits from the entropy of the corresponding motion and residual coefficients.

\section{Procedure}
\label{sec_Optimization Procedure}

\subsection{Offline Hybrid Optimization Procedure}
\label{sec_procedure_offline}
Based on those designed modules, we build our complete coding framework.
Our framework is on top of deep networks, so it models the optimization problem with a continuous function.
Then we end-to-end train the framework with gradient-based optimization through lots of samples. The details of the network training will be provided in Section \ref{sec_Network Training}.

Meanwhile, our framework contains multiple modes.
The variable block partition, multiple motion modes, and multiple residual modes can be combined so that it will produce a large number of discrete starting points for search optimization.
We use the trained model parameters to initialize those modes.
Specifically, the partition mode is initialized to four sizes, and the residual coding mode is initialized to whether regular transform coding or the zero residual.
As for motion modes, all blocks follow the same initialization rules.
The coefficients of temporal merge mode are initialized to the fixed value $0.5$, and the scale coefficients of temporal scale mode are initialized by
\begin{equation}
s_{0} = \frac{\overline{v}_{t \rightarrow t-k}}{\overline{v}_{t+k \rightarrow t-k}}, s_{1} = \frac{\overline{v}_{t \rightarrow t+k}}{\overline{v}_{t-k \rightarrow t+k}}
\label{equ_TScale_Init}
\end{equation}
and the MVs of motion vector mode are initialized by
\begin{equation}
mv_{t \rightarrow t-k} = \overline{v}_{t \rightarrow t-k}, mv_{t \rightarrow t+k} = \overline{v}_{t \rightarrow t+k}
\label{equ_MV_Init}
\end{equation}
where $\overline{v}$ means to calculate the average of the optical flow in a block.
As a result, we finish the offline optimization of the coding framework through lots of samples.

\subsection{Online Hybrid Optimization Procedure}
\label{sec_procedure_online}

\begin{algorithm}
\caption{Online Hybrid-Optimization Process}
\label{alg1}
    \begin{algorithmic}[1]
    \Require
            $ME()$: motion estimation, which inputs original frame ($orig$) and reference frames ($ref$) and outputs optical flow ($flow$);
            $MC()$: motion compensation, which inputs motion coefficients ($c$) and reference frames ($ref$) and outputs predicted frame ($pred$);
            $CC()$: motion coefficients coding, which inputs motion coefficients and outputs coefficient bits map ($cbm$);
            $RC()$: residual coding, which inputs residual frame and outputs reconstructed frame ($reco$) and residual bits map ($rbm$);
            Note that those modules are all operated at frame level;
    \State $flow = ME(orig, ref)$;
    \For{each depth $d\in [0,D)$}
        \State Partition frame into blocks with size $z_d=z_0<<d$;
        \For{each motion mode $l\in [0,L)$}
            \For{each block $n\in [0,N_d)$ in frame}
                \State Initialize $c_{d,l}^n$ with $flow$ for each block;
            \EndFor
            \State $pred_{d,l} = MC(c_{d,l}, ref)$, $cbm_{d,l} = CC(c_{d,l})$;
            \State $reco_{d,l}, rbm_{d,l} = RC(orig - pred_{d,l})$;
            \State Calculate RD cost $J_{d,l}$ for each block;
            \If{$c_{d,l}$ need online numerical optimization}
                \State Optimize $c_{d,l}$ online by minimizing $J_{d,l}$;
            \EndIf
        \EndFor
        \For{each block $n\in [0,N_d)$}
            \State Search motion mode $J_d^n = min\{J_{d,l}^n, l \in [0,L)\}$;
        \EndFor
        \State Update $J_d$ by the selected mode coefficients $c_d$;
        \If{depth $d == 0$}
            \State Take $J_d$ as best RD cost until depth $d$, i.e.$\widetilde{J}_d = J_d$;
        \Else
            \For{each block $n\in [0,N_d)$}
                \State Search partition $\widetilde{J}_d^n = min\{\sum_{i=1}^4\widetilde{J}_{d-1}^{n_i}, J_d^n\}$;
            \EndFor
            \State Update $\widetilde{J}_d$ by the selected partition and mode;
        \EndIf
    \EndFor
    \State Obtain the best RD cost $\widetilde{J}=\widetilde{J}_d$;
    \State Calculate RD cost of skipping residual $\hat{J}$ for each unit;
    \For{each unit $u\in [0,U)$}
        \State Search residual mode $J^u = min\{\hat{J}^u, \widetilde{J}^u\}$;
    \EndFor
    \State Update $J$ by the selected residual mode;
    \State Code final partition, mode, motion\&residual coefficients;
    \end{algorithmic}
    \label{algorithmic_coding_process}
\end{algorithm}

Based on our offline optimized framework, we conduct online hybrid optimization for each frame at the encoder to find the better coding parameters. The optimization framework is shown in Fig.\ref{fig Video_Compression_Framework} (b), and the optimization process is shown in Algorithm \ref{algorithmic_coding_process}. In this paper, we set the number of partition depth $D = 4$, minimum partition block size $z_0 = 8$, and the number of motion mode $L = 3$, which have already been introduced in the above sections.
As the proposed online hybrid-optimization method in Section \ref{sec_method_online}, we perform continuous optimization in local parts and discrete optimization in global space in turn.
Firstly, in the local area of each specific mode, we perform continuous numerical optimization around the initial mode parameters.
Specifically, we traverse the block partition first, and then we optimize the parameters of each motion mode separately under a specific partition.
The coefficients of temporal merge mode are not coded, i.e. fixed value $0.5$, so they don't need to be optimized. For other motion modes, we get the initial coefficients for each block firstly, where the initialization rules have been introduced in the offline optimization procedure of Section \ref{sec_procedure_offline}.
Since not all frames have significant coding gains through online gradient-based optimization, such as the frame with little or no motion, we optimize some frames selectively in a video sequence.
If the current frame needs to be optimized, we online end-to-end optimize all coefficients of the current mode at the frame level with gradient-based optimization by minimizing the RD cost $J$:
\begin{equation}
J = D + \lambda R = d(\hat{x}, x) + \lambda(R_{m} + R_{r})
\label{equ_RD_detail}
\end{equation}
where $d(\hat{x}, x)$ denotes the distortion between original frame $x$ and reconstructed frame $\hat{x}$, and we use sum square error (SSE) in our implementation. $R_{m}$ and $R_{r}$ represent the number of bits for motion information and residual in this frame, respectively.
$\lambda$ of each frame in our framework is set to be the same as HM.
The iteration number is set to 10.

After obtaining the local optimal coefficients of all modes in the current partition depth through continuous numerical optimization, we perform discrete search optimization among them, i.e. selecting the best mode for each block according to the RD cost.
The block's RD cost is calculated with the distortion between the original block and the reconstructed block and the bits of motion and residual in this block.
Note that the networks in our framework are all operated at the frame level. Thus, the distortion and bits of each block are estimated from the frame result.
After all search optimization at the block level, the reconstructed frame needs to be updated at the frame level through the networks instead of block splicing.
Then we have finished the motion mode optimization in the current partition depth.

Subsequently, we perform discrete search optimization for the partition mode of each region by comparing the RD cost of the block in the current depth with the best result of this region before the current depth.
We repeat that discrete optimization until finishing all partition modes.
In the end, we optimize the residual skip mode for each region by discrete search.
As a result, based on the offline model, through the continuous numerical optimization around the starting points locally and the discrete search optimization among all local optimums globally, we have optimized the better parameters for the coded sample in a hybrid manner.

\section{Experiments}
\label{sec_Experimental_Results}

\subsection{Experimental Settings}
\label{sec_Experimental_Settings}

\subsubsection{Network Training}
\label{sec_Network Training}
We use the deep learning software PyTorch to train our framework on 4 NVIDIA GTX 1080Ti GPUs.
When training our networks, we remove all modes and code the pixel-level optical flow between the current frame and reference frames directly so that the whole network can be trained in an end-to-end manner.
The goal of video compression is to optimize rate-distortion, so we take Eq. \eqref{equ_RD_detail} as the loss function to minimize the RD cost.
We train 4 basic models for 4 common test QPs in HM.
We use $\lambda$ and transform quantization step $Q_{step}$ to control the bitrate of different models.
In our framework, we set $Q_{step}$ to 12, 24, 45, 95 to correspond to QP 22, 27, 32, 37 in HM and set $\lambda$ to be the same as HM.
Since the quantization operation is not differential, we adopt the method of forward quantization and backward pass \cite{theis2017lossy}.
Each group of training data consists of three frames, including the original frame, the reference frame before and after the original frame. The reference frames should be reconstructed from our compression network, but our network has not yet completed training. To address this problem, we compress the reference frames by HM under four QPs to approximate them.
We use the SJTU-4k dataset \cite{song2013sjtu} and down-sampling versions of these video sequences.
We extract only the luma component for training. We directly use the luma models to code the chroma in the test.
The frame is cropped into many non-overlapping $128\times128$ blocks, and the time interval between the original frame and reference frames is set to 1 or 2 randomly to fit the hierarchical layers in the test.
Finally, we obtain about 500,000 training samples for each model.
We optimize with Adam \cite{kingma2014adam} using 0.9 and 0.999 as hyper-parameters with no weight decay. The learning rate starts at 0.0001 and then is reduced to 0.00001 when the loss becomes stable. The batch size is 16.

\subsubsection{Encoding Configurations and Evaluation}
\label{sec_Encoding Configurations and Evaluation}
We evaluate our coding framework on the HEVC common test sequences, including 16 videos of different resolutions known as Classes B, C, D, E \cite{bossen2011common}.
Considering that the luma component (Y) is more important than the chroma components (U and V), we provide coding results of test sequences with two formats: YUV400 and YUV420.
The YUV400 is the default format in the following experiments.
We compress the first 1 second of each sequence when comparing our proposal to HEVC. We compress the first 100 frames of each sequence when comparing our proposal to end-to-end learned methods.
PSNR is used to measure the quality of the reconstructed frames, and BD-rate \cite{bjontegaard2001calcuation} is calculated to quantify the coding performance between different schemes.
We compare our coding framework with the vanilla HEVC reference software HM (version 16.10).
We follow the HEVC common test condition and use the default Random-Access (RA) encoding configurations provided in HM16.10 without any change, if not otherwise specified. HM is tested under 4 QPs: 22, 27, 32, and 37 \cite{bossen2011common}.
Our framework is tested with 4 trained $Q_{step}$ models: 12, 24, 45, and 95. In a basic model, we can adjust $Q_{step}$ slightly to easily achieve the hierarchical quality of different layers.
The intra period (IP) of HM and our framework are both 1 second.
The GOP of HM and our framework are both 8.
HM coding is running in CPU and our framework coding is running in CPU and a single GPU. The CPU is Intel(R) Xeon(R) CPU E5-2690 v4 @2.60GHz and GPU is NVIDIA TITAN Xp with 12 GB RAM.

\subsection{Overall Performance}
\label{sec_Overall Performance}

\begin{table}
\centering
\caption{BD-rate results of our proposal compared to HM}
\begin{tabular}{l|l|r}
 \hline
 \multirow{2}{*}{Class}& \multirow{2}{*}{Sequence}&\multicolumn{1}{c}{RA}\\
 &&Y (\%)\\
 \hline
 \multirow{5}{*}{Class B}
& Kimono                           & $-$2.4       \\
&ParkScene                       &$-$4.4           \\
 &Cactus                         &$-$4.1           \\
&BasketballDrive                 &2.5           \\
&BQTerrace                       &43.4           \\
 \hline
\multirow{4}{*}{Class C}
& BasketballDrill                &2.6          \\
 &BQMall                         &4.9         \\
&PartyScene                      &0.4         \\
&RaceHorsesC                      &16.6         \\
 \hline
 \multirow{4}{*}{Class D}
 & BasketballPass                 &$-$6.4       \\
&BQSquare                         &$-$14.5       \\
 &BlowingBubbles                  &$-$0.8        \\
&RaceHorses                       &3.9        \\
 \hline
\multirow{3}{*}{Class E}
& FourPeople                      &$-$4.6      \\
 &Johnny                          &2.9      \\
&KristenAndSara                   &1.2       \\
\hline
 \multirow{4}{*}{Class Summary}
 &Class B                        &7.0    \\
 &Class C                        &6.1    \\
 &Class D                        &$-$4.4  \\
 &Class E                        &$-$0.2\\
 \hline
\textbf{Overall}    &\textbf{Classes B--E}                & \textbf{2.6} \\
 \hline
\end{tabular}
\label{table_performance proposal}
\end{table}

\begin{table}
\centering
\caption{BD-rate results of our proposal compared to HM on YUV420}
\begin{tabular}{l|l|rrr}
 \hline
 \multirow{2}{*}{Class}& \multirow{2}{*}{Sequence}&\multicolumn{3}{c}{RA}\\
 &&Y (\%) &U (\%) &V (\%)\\
 \hline
 \multirow{5}{*}{Class B}
& Kimono                           & 0.9   &0.4  & 1.0    \\
&ParkScene                       &$-$3.5   &1.3  &2.0        \\
 &Cactus                         &$-$2.9   &$-$2.5  &$-$0.7        \\
&BasketballDrive                 &4.3   &12.0  &5.1          \\
&BQTerrace                       &44.9   &10.5  &11.2          \\
 \hline
\multirow{4}{*}{Class C}
& BasketballDrill                &4.2    &12.9  &9.0        \\
 &BQMall                         &6.2     &13.5  &11.1      \\
&PartyScene                      &0.6     &3.4  &3.7      \\
&RaceHorsesC                      &17.3      &28.1  &24.6     \\
 \hline
 \multirow{4}{*}{Class D}
 & BasketballPass                 &$-$4.4    &3.2  &$-$4.0     \\
&BQSquare                         &$-$13.6     &$-$2.9  &$-$6.2    \\
 &BlowingBubbles                  &$-$0.5      &3.6  &5.5    \\
&RaceHorses                       &6.8       &21.2  &14.2   \\
 \hline
\multirow{3}{*}{Class E}
& FourPeople                      &$-$4.4     &$-$2.8  &$-$2.9   \\
 &Johnny                          &2.7     &0.2  &2.2   \\
&KristenAndSara                   &1.3      &$-$0.6  &$-$0.5   \\
\hline
 \multirow{4}{*}{Summary}
 &Class B                        &8.7    &4.3  &3.7  \\
 &Class C                        &7.1     &14.5  &12.1 \\
 &Class D                        &$-$2.9    &6.3  &2.4\\
 &Class E                        &$-$0.1   &$-$1.1  &$-$0.4\\
 \hline
\textbf{Overall} &\textbf{Classes B--E} &\textbf{3.7} &\textbf{6.3} &\textbf{4.7} \\
 \hline
\end{tabular}
\label{table_performance proposal_YUV420}
\end{table}

\begin{figure}
  \centering
  \includegraphics[width=\columnwidth]{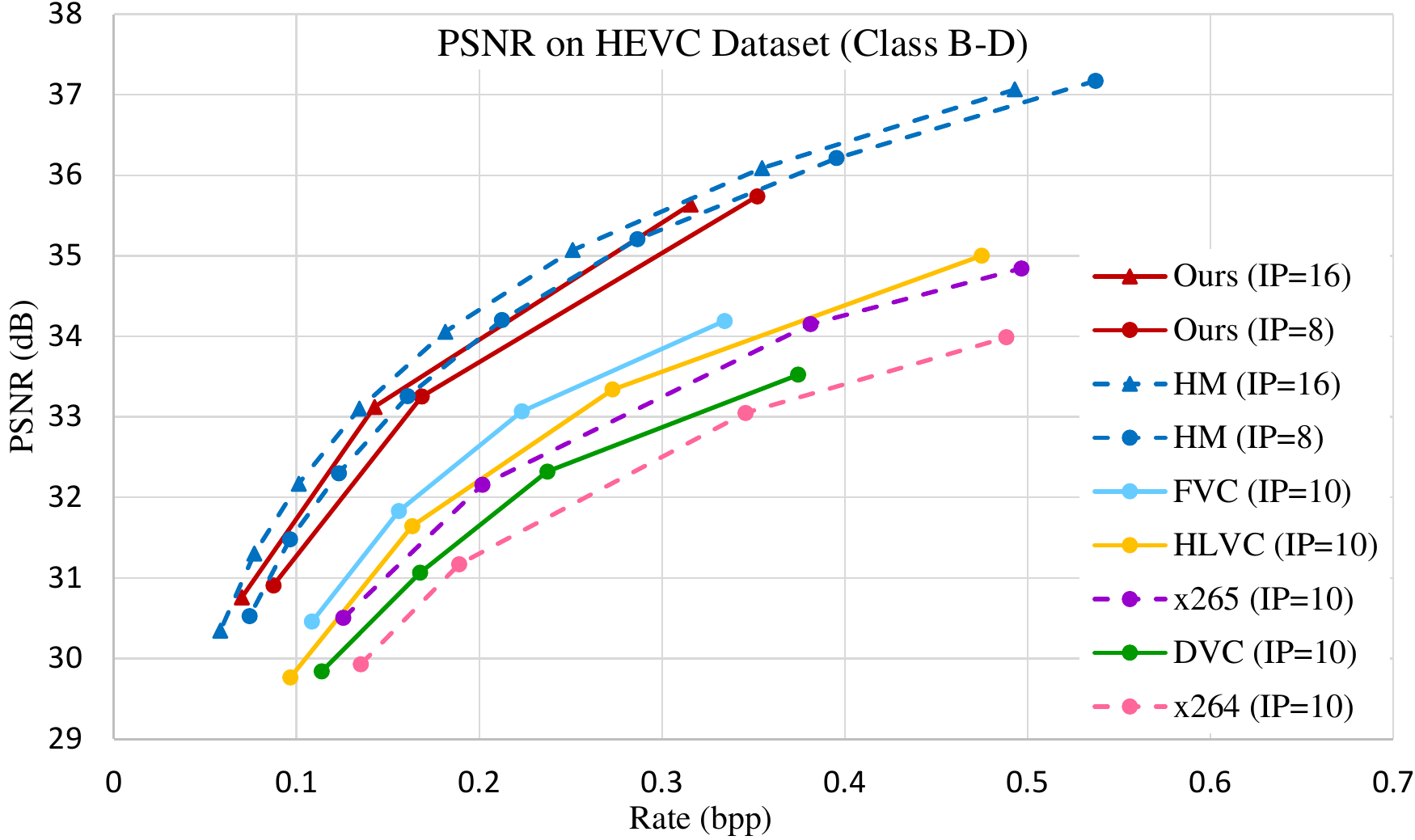}
  \caption{Rate-distortion comparison in terms of PSNR on HEVC dataset (Classes B--D) using x264, x265, DVC\cite{lu2019dvc}, HLVC\cite{yang2020learning}, FVC\cite{hu2021fvc}, HM, and our proposed method. The intra period(IP) of x264, x265, DVC, HLVC, and FVC reported in \cite{lu2019dvc, yang2020learning,hu2021fvc} is 10. For a fair comparison, the IP of HM and our proposed method is set to 8 and 16, respectively.}
  \label{fig Other_Methods_Comparison_Curve}
\end{figure}

To verify the efficiency of our proposed hybrid optimization, we compare our framework with both block-based hybrid video coding and end-to-end learned video coding.
The basic ideas of all designed modes in our framework come from HEVC, i.e. quadtree partition, temporal merge and scale mode, motion vector mode, and residual skip mode in our framework corresponding to quadtree partition, temporal merge and AMVP candidate, motion vector mode, and skip mode in HEVC.
Therefore, we choose the HEVC of block-based hybrid video coding for comparison.
The BD-rate results of our framework compared to HEVC reference software HM16.10 in YUV400 format are shown in Table \ref{table_performance proposal}. It can be observed that our proposed hybrid-optimization video coding method achieves comparable performance with HM in terms of PSNR, leading to on average only 2.6\% BD-rate increase on Y component. Our method achieves better performance on half of the classes and beats HM on 7 sequences among all 16 test sequences.
In addition, we provide the BD-rate comparison results of YUV420 format in Table \ref{table_performance proposal_YUV420}. Our proposal achieves on average 3.7\%, 6.3\%, and 4.7\% BD-rate increase for Y, U, and V, respectively. The performance on the two formats is consistent.
Our method doesn't completely outperform HEVC mainly because our current framework only uses a very little part of modes in HEVC. In addition to those used modes, HEVC contains many other advanced modes, such as intra mode, non-square PU mode, spatial merge and AMVP candidate, uni/bi-prediction mode, various transform modes, RDOQ, etc. Besides, HEVC also uses additional information, such as more reference frames.
Therefore, if we add more modes or information into our framework, it will have the potential to outperform HEVC.

The learned end-to-end video coding methods usually code the video in RGB format.
To compare with them, our framework codes the video of YUV420 format firstly and then converts the reconstruction and the original video into RGB format to calculate RGB PSNR.
Fig.\ref{fig Other_Methods_Comparison_Curve} shows the average performance of some methods in RGB space on the HEVC dataset (Classes B, C, D), including the typical learned method DVC\cite{lu2019dvc}, the latest bi-directional learned method HLVC\cite{yang2020learning}, the state-of-the-art learned method FVC\cite{hu2021fvc}, x264, x265, HM under RA configuration and our method.
DVC, HLVC, and FVC only provide the results with the intra period (IP) 10. Since the IP in HM needs to be set to a multiple of GOP size (GOP size is 8 in HM16.10) and our framework adopts a similar hierarchical structure with HM, we provide the performance of HM and our method with IP 8 and IP 16. In theory, the performance of IP 10 should be between IP 8 and IP 16. It can be observed that the coding performance of our hybrid-optimization coding method is significantly better than the end-to-end learned methods, leading to more than 45\% bits saving compared to the HLVC and more than 30\% bits saving compared to the state-of-the-art FVC.

\subsection{Verification of Hybrid Optimization}
\label{sec_Ablation Study}
\begin{table*}
\renewcommand\arraystretch{1.3}
\centering
\caption{Ablation study: BD-rate results of our proposed modes. Anchor is HM. (1) E2E TMerge: our end-to-end scheme using temporal merge mode. (2) E2E Flow: our end-to-end scheme using pixel-level flow. (3) + TScale: add temporal scale mode with fixed block partition ($32\times32$) to (1). (4) + MV: add motion vector mode with fixed block partition ($32\times32$) to (3). (5) + VBlock: add variable block partition to (4). (6) + ResiSkip: add residual skip mode to (5).}
\begin{tabular}{c|c|c|c|c|c|c}
 \hline
 &{E2E TMerge} &{E2E Flow} &{+ TScale} &{+ MV}  &{+ VBlock}   &{+ ResiSkip}       \\
\cline{2-7}
 &Y (\%) & Y (\%) &Y (\%) &Y (\%) & Y (\%) &Y (\%)\\
\hline
Class B &29.4  &130.1       &19.9       &13.3     &8.2            &\textbf{8.3}  \\
\hline
Class C &21.6  &66.5       &15.6       &10.0     &7.0             &\textbf{6.8}   \\
\hline
Class D &7.6  &33.5        &2.8        &$-$0.4   &$-$2.9          &\textbf{$-$3.5}      \\
\hline
\textbf{Average} &20.3 &80.8 &13.3    &8.0       &4.4             &\textbf{4.2}  \\
\hline
\end{tabular}
\label{table_perfomance_ablation_study}
\end{table*}

\begin{table}
\centering
\caption{BD-rate results of our proposal with online gradient-based optimization compared to our proposal without online gradient-based optimization on the Layer 1 frames}
\begin{tabular}{l|l|c}
 \hline
& Sequence &Y (\%)\\
 \hline
 \multirow{5}{*}{Class B}
& Kimono                         &$-$8.8       \\
&ParkScene                       &$-$0.8           \\
&Cactus                          &$-$1.1           \\
&BasketballDrive                 &$-$6.5           \\
&BQTerrace                       &$-$7.7           \\
 \hline
\multirow{4}{*}{Class C}
& BasketballDrill                &$-$0.5          \\
 &BQMall                         &$-$6.9         \\
&PartyScene                      &$-$2.5         \\
&RaceHorsesC                     &$-$3.1         \\
 \hline
 \multirow{4}{*}{Class D}
& BasketballPass                 &$-$4.1       \\
&BQSquare                        &$-$2.8       \\
&BlowingBubbles                  &$-$3.6        \\
&RaceHorses                      &$-$5.4        \\
 \hline
 \textbf{Overall}    &\textbf{Classes B--D}                & \textbf{$-$4.1} \\
  \hline
\end{tabular}
\label{table_performance OnlineTraining}
\end{table}

In order to further verify the effectiveness of each optimization in our proposed hybrid-optimization coding method, we perform ablation studies on Classes B, C, D.
First, we take the popular end-to-end learned method as the baseline, i.e. offline end-to-end gradient-based optimization and online direct inference, and provide two schemes. The one is called E2E TMerge, which obtains the predicted frame by interpolating from reference frames without transmitting any motion parameters and then codes the residual frame. The other one is called E2E Flow, which obtains the predicted frame by directly transmitting pixel-level optical flow and then codes the residual frame. For a fair comparison, we code the pixel-level flow with the same method as the MV in MV mode, i.e. coding the vector directly without transform.
The BD-rate results are shown in Table \ref{table_perfomance_ablation_study}. It can be observed that the E2E TMerge performs better than the E2E Flow, leading to on average 20.3\% BD-rate increase compared to HM.
It's mainly because it's not efficient to code pixel-level motion directly in hierarchical bi-directional coding structure.
Next, based on the E2E TMerge, we add discrete multi-mode search-based optimization gradually and conduct
a comprehensive set of experiments to verify the effectiveness of each proposed mode. As shown in Table \ref{table_perfomance_ablation_study}, when gradually adding TScale mode, MV mode, variable block partition mode, and ResiSkip mode, the coding performance improves significantly and the BD-rate increase decreases to 13.3\%, 8.0\%, 4.4\%, and 4.2\%, respectively. The results show that our method outperforms the pure end-to-end learned coding method and all proposed mode are effective. More importantly, it further proves that it is efficient to mix discrete multi-mode search-based optimization into continuous end-to-end gradient-based optimization.
In addition to those optimization methods, our framework also contains online continuous gradient-based optimization in local parts, and we conduct experiments to verify its effectiveness. Since some frames in hierarchical layer structure may have very small and simple motion, such as the high layers with short time intervals, it is not necessary to online optimize all frames. We only perform online gradient-based optimization on the frames of layer 1. The BD-rate results are shown in Table \ref{table_performance OnlineTraining}, where we only calculate the bits and PSNR of the layer 1 frames. We can observe that the online continuous gradient-based optimization improves the coding efficiency significantly, leading to on average 4.1\% BD-rate reduction. Especially, it reaches very high coding gain on some sequences, such as 8.8\% in \texttt{Kimono} and 7.7\% in \texttt{BQTerrace}. Through these experiments, we fully verified the effectiveness of the proposed hybrid optimization.

\subsection{Detailed Analysis}
\label{sec_Analysis}
\subsubsection{Performance Analysis}
\label{sec_Performance Analysis}

In particular, our proposal achieves BD-rate reduction in Classes D and E, leading to on average 4.4\%, 0.2\%, respectively. Class D, a class with low-resolution sequences, has a small moving range of pixels correspondingly, and Class E almost has no motion. Those sequences can be coded efficiently because the networks can deal with small motion easily. Especially for the sequence \texttt{BQSquare}, the bits saving is up to 14.5\%.
On the other hand, our proposal has a slight BD-rate increase on average in Classes B and C.
It's mainly due to \texttt{BQTerrace} and \texttt{RaceHorsesC}.
Except for the two sequences, our proposal achieves even better performance on average than HM.
\texttt{BQTerrace} features lots of noise, aliasing, and textures, including dynamic textures on the water surfaces and the repeated texture on the wall surface and railings.
\texttt{RaceHorsesC} contains the entangled occlusion on the fast moving horses and the texture on the grass.
Texture, complex object motion, large motion, and entangled occlusion have always been the huge challenge for optical flow estimation, and network-based optical flow estimation methods \cite{teed2020raft,yin2019hierarchical} also have not well solved those problems so far.
Our method uses optical flow as the initial motion information to conduct online gradient-based optimization. When the initial optical flow is inaccurate, it is difficult for our online gradient-based optimization to obtain precise motion information within few optimization times.
Consequently, our proposal leads to worse performance on \texttt{BQTerrace} and \texttt{RaceHorsesC}.

\subsubsection{Mode Selection Analysis}
\label{sec_Mode Selection Analysis}
\begin{figure*}
  \centering
\includegraphics[width=\linewidth]{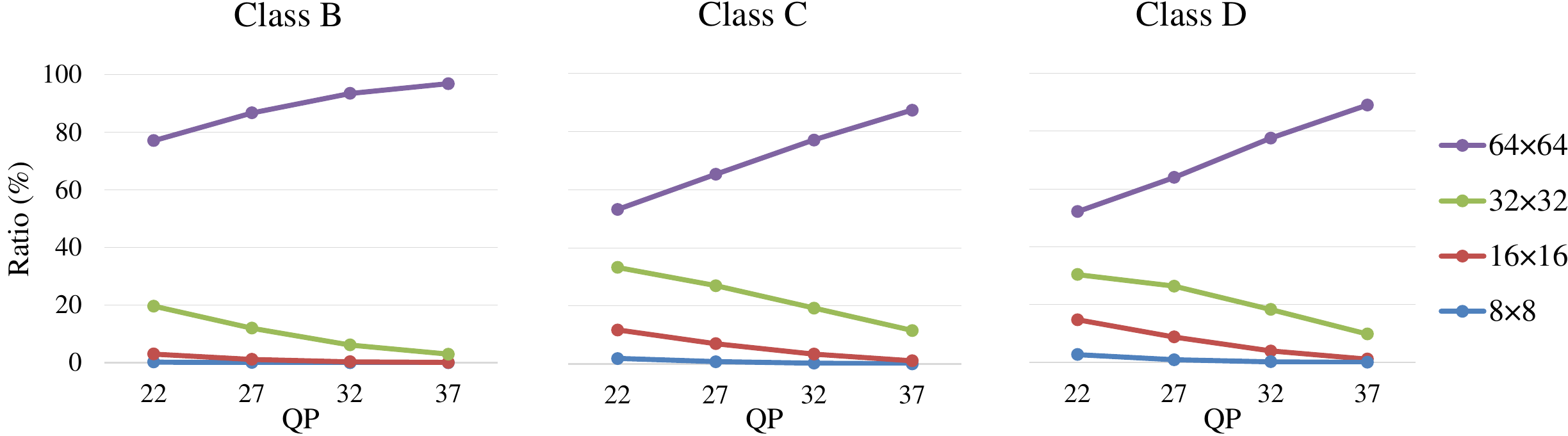}
\caption{Partition selection ratio results.}
\label{fig PartitionSelectionRatio}
\end{figure*}
\begin{figure*}
  \centering
\includegraphics[width=\linewidth]{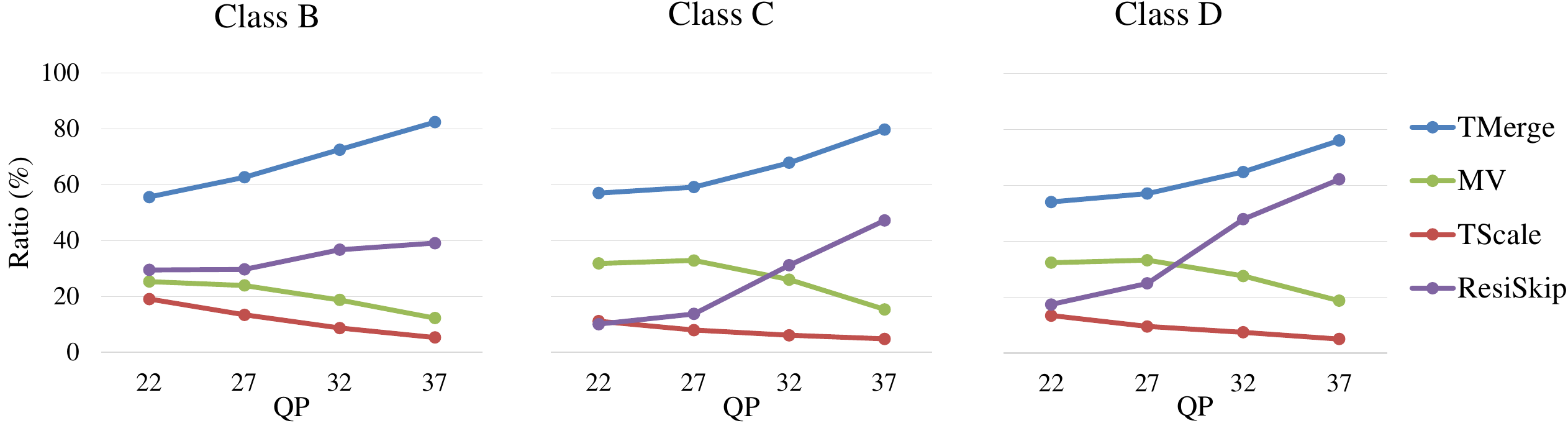}
\caption{Mode selection ratio results.}
\label{fig ModeSelectionRatio}
\end{figure*}
\begin{figure*}
\centering
\subfigure[ ]
{
\includegraphics[width=0.45\linewidth]{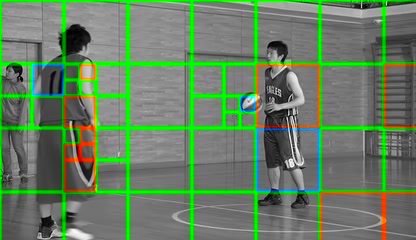}
}
\subfigure[ ]
{
\includegraphics[width=0.45\linewidth]{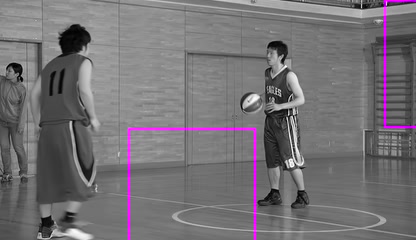}
}
\subfigure[ ]
{
\includegraphics[width=0.45\linewidth]{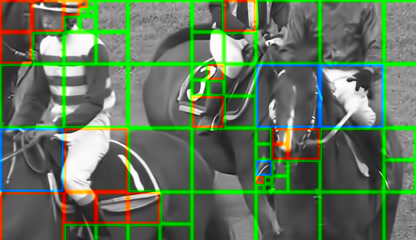}
}
\subfigure[ ]
{
\includegraphics[width=0.45\linewidth]{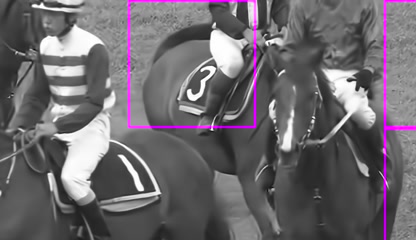}
}
\caption{Left: visual results of block partition and motion mode selection. Green, blue, red indicates blocks that choose Temporal Merge Mode, Temporal Scale Mode, Motion Vector Mode, respectively. Right: visual results of residual mode selection. Purple indicates blocks that choose Residual Skip Mode. Top: 7-th frame of BasketballPass at QP 22. Bottom: 21-st frame of RaceHorses at QP 32.}
\label{fig_visual_mode_selection}
\end{figure*}
\begin{figure}
  \centering
  \includegraphics[width=\columnwidth]{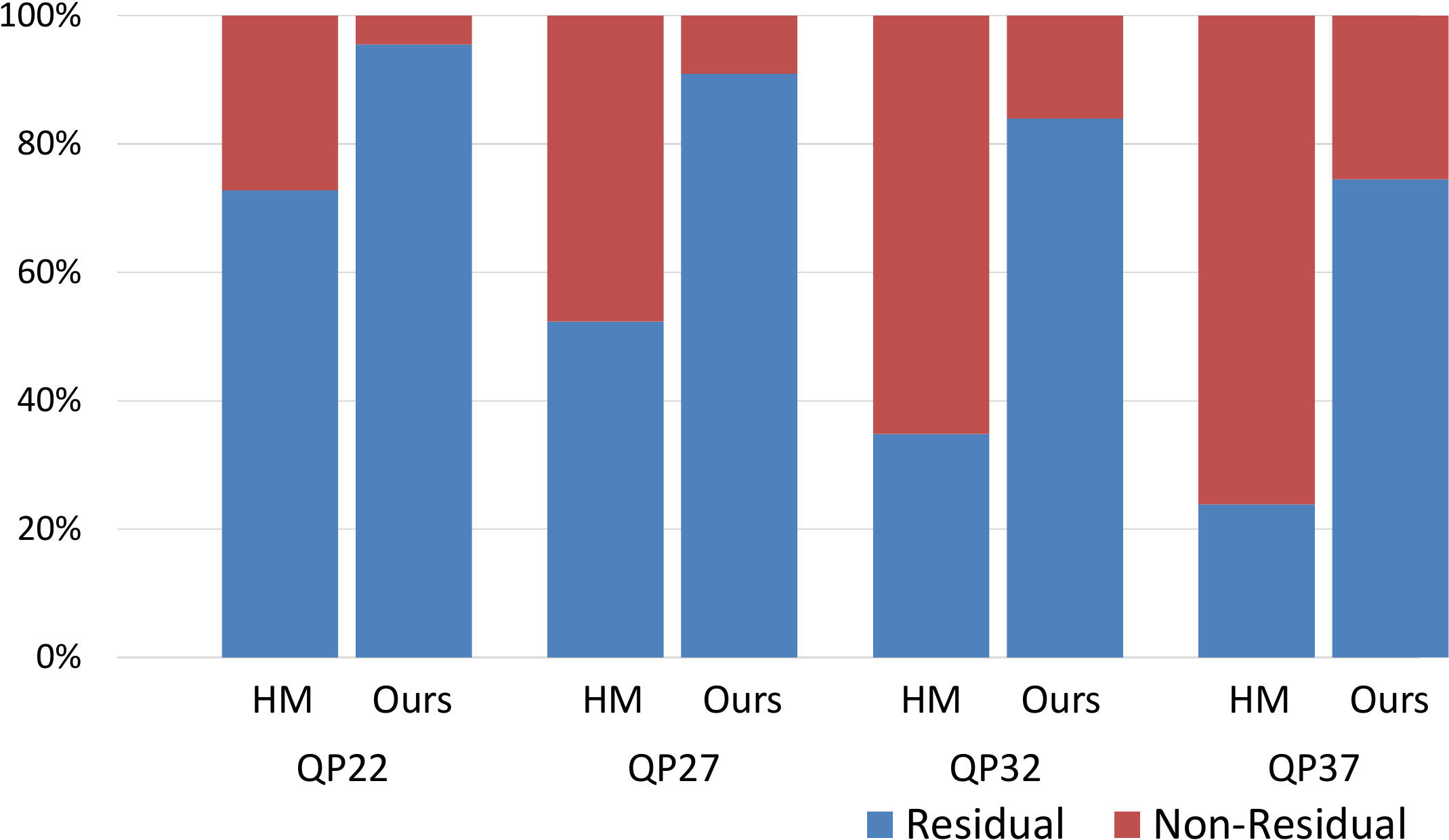}
  \caption{Comparison of the proportion of each component in the bitstream between HM and our proposal.}
  \label{fig Proportion_of_Residual_Bit}
\end{figure}

We analyze the results of mode selection to gain some insights. Particularly, we perform statistics of the hitting ratio, defined as the ratio of blocks choosing a mode over all the partitioned blocks. Here the ratio is calculated by area rather than by count because blocks have different sizes in our framework.
For the block partition mode, the hitting ratio statistics are shown in Fig. \ref{fig PartitionSelectionRatio}.
First, the larger the block, the higher the hitting ratio.
Second, as QP increases, the hitting ratio of the largest block $64\times64$ increases, and the hitting ratio of other small blocks, i.e. $8\times8$, $16\times16$, and $32\times32$, decreases.
Third, as resolution increases, the hitting ratio of the largest block increases, and the hitting ratio of other small blocks decreases.
One of the reasons for these phenomenons is that larger block partition can reduce overhead, especially at lower bit rates.
In addition, since the motion of pixels has a strong spatial correlation, especially in high resolution, a large block partition can utilize the correlation more efficiently.

For the motion and residual modes, the hitting ratio statistics are shown in Fig. \ref{fig ModeSelectionRatio}.
First, the motion modes in descending order of the hitting ratio are TMerge, MV, TScale. It's mainly because a small part of regions can fully meet the requirements of TScale, and motion in most regions can be considered uniform in such a short time interval and then choose TMerge.
Second, as QP increases, the hitting ratio of TMerge and ResiSkip increases, and the hitting ratio of MV and TScale decreases. The major reason is that TMerge and ResiSkip can save the bits of motion parameter and residual significantly, which is important for lower bit rates.
Third, the motion and residual modes are less related to resolution, because the motion and residual mainly depend on the video content.
In summary, the large partitioned block, TMerge mode, and ResiSkip mode can reduce overhead, which is beneficial for lower bit rates. The small partitioned block, TScale mode, and MV mode can provide a fine description for motion, which is efficient for higher bit rates.
Therefore, there is no remarkable performance difference between higher and lower bit rates, as shown in Fig.\ref{fig Other_Methods_Comparison_Curve}.

Fig. \ref{fig_visual_mode_selection} presents some visual results of mode selection.
We can observe that large blocks, TMerge mode, ResiSkip mode are selected for the background and the region with simple motion, like the floor in Fig. \ref{fig_visual_mode_selection} (a) and (b) and the grassland, horse bodies in Fig. \ref{fig_visual_mode_selection} (c) and (d), whilst small blocks are selected for the region with complex motion and the edge, like the moving basketball players in Fig. \ref{fig_visual_mode_selection} (a), the edge of riders, horse heads and legs in Fig. \ref{fig_visual_mode_selection} (c). TScale mode is selected for the region with regular temporally non-uniform motion, like the basketball with accelerated rotation, the legs with the accelerated movement of the right-side player in Fig. \ref{fig_visual_mode_selection} (a), the horse head from rising to suddenly bowing in Fig. \ref{fig_visual_mode_selection} (c). MV mode is selected for the translational region and the region with the inaccurate prior flow of reference frames, especially in the picture boundary and the occlusion, like the player in Fig. \ref{fig_visual_mode_selection} (a), the rider legs in Fig. \ref{fig_visual_mode_selection} (c).
Those visual results are consistent with our design.

Moreover, we also perform statistics of the inter-prediction and residual bits in the bitstream between HM and our proposal. The ratio results are shown in Fig. \ref{fig Proportion_of_Residual_Bit}. First, the ratio of residual bits increases in both our proposal and HM as bit rate increases. Second, the ratio of residual bits in our proposal is higher than HM. The main reason is that our TMerge mode can provide an accurate inter-prediction without any bits, and most regions choose TMerge mode as shown in Fig. \ref{fig ModeSelectionRatio}. Thus, the ratio of inter-prediction bits in our proposal becomes lower, and the ratio of residual bits will be higher correspondingly.

\subsection{Computational Complexity}
\label{sec_Computational Complexity}
\begin{table*}
\centering
\caption{Average running time per frame of our proposal and HM for a Class D sequence (BQSquare $416\times240$). QP is 27. (1) HM. (2) Our proposal, including the detailed time of each module.}
\begin{tabular}{l|c|cccccccc}
\hline
\multirow{2}{*}{Time (s)} & \multirow{2}{*}{HM}    & \multicolumn{8}{c}{Proposal}  \\
\cline{3-10}
\multirow{2}{*}{}&\multirow{2}{*}{} &Optical Flow   &MC    & Transform\&Quantization  &Context Model &AE/AD  &In-Loop Filter &Mode Decision  &Total\\
\hline
Enc.                  &2.29         &0.33     &0.056     &0.014        &0.835          &0.809        &0.006                &6.143         &\textbf{8.193}         \\
\hline
Dec.                  &0.004        &0.168    &0.056     &0.008        &86.125         &0.799        &0.006                &$-$            &\textbf{87.162}  \\
\hline
\end{tabular}
\label{computational_time}
\end{table*}

We record the encoding and decoding time of our proposed framework and the vanilla HM.
We take a Class D sequence (\texttt{BQSquare} $416\times240$) as an example and the computational time results are shown in Table \ref{computational_time}. Compared to HM, the encoder time of our framework increases slightly while the decoder time increases significantly. In order to know where the computational complexity comes from, we provide the detailed results of each module in our framework to gain some insights. It can be observed from Table \ref{computational_time} that the optical flow, MC, transform and quantization, and in-loop filter module are fast.
The computational time mainly comes from the mode decision optimization at the encoder and the context model at the decoder.
At the encoder side, we compare multiple modes and update the compared results many times through the networks, which takes lots of computational time.
As for the context model, we can estimate the probability distribution for all coefficients in parallel at the encoder side. However, at the decoder side, we must deal with each coefficient sequentially because the probability distribution of the current coefficient needs to be based on the previous coefficients.
It is important to note that if designing a customized algorithm for context model in deep learning software or using dedicated hardware for CNN inference, the time can be further reduced. Future work is required to make the proposed method suitable for real-time applications.

\section{Discussion}
\label{sec_Discussion}
In addition to rate and distortion, video coding also needs to consider coding complexity in practical applications. Accordingly, the potential of a video coding framework is mainly judged by two aspects: rate-distortion performance and computational complexity. Next, we will discuss our hybrid optimization from the two aspects to show its advantages and potential.

In the past, researchers study advanced coding methods on block-based hybrid video coding and end-to-end learned video coding independently. However, due to the gap between the two frameworks, the advanced schemes of one framework often can't be applied to the other framework.
As a result, those advanced technologies from the two frameworks can't improve coding performance together, which severely limits the development of video coding.
Our hybrid-optimization coding framework as the bridge between the two frameworks can solve this problem well.
Since the coding technology is essentially based on optimization methods, when we carry out the hybrid in optimization, those technologies can be combined accordingly.
Specifically, better models in end-to-end learned video coding, such as optical flow estimation, bi-directional prediction, entropy coding, and in-loop filter, can be easily plugged into our hybrid-optimization framework to improve coding performance.
In addition, the multi-mode comparison optimization makes our framework easy to add the more advanced modes from block-based hybrid video coding to improve coding performance, such as intra mode, multi-type tree partition, affine mode, merge mode, uni/bi-prediction mode, RDOQ, and so on.
Our hybrid optimization framework can attract and combine most of the new technologies from the two frameworks.
Therefore, our hybrid optimization has significant potential to improve RD performance.

Generally, the search-based optimization in the block-based hybrid video coding is performed on the CPU, while the numerical optimization in end-to-end learned video coding is carried out on the GPU.
In practical applications, the devices that perform video coding by software usually have both CPU and GPU, such as the widely used modern personal computers (PC).
However, the two frameworks usually only use one of the CPU and GPU, wasting the computing resources of the other one.
On the other hand, the hybrid of the two optimizations will lead to the hybrid in the computing architecture.
Our hybrid optimization can adopt distributed computing to perform different optimization on different computing devices so that both CPU and GPU can be fully utilized at the same time. Therefore, hybrid optimization has great potential in future practical applications.

\section{Conclusion}
\label{sec_Conclusion}
In this paper, we no longer look at video coding from the technical point of view, but rethink video coding from its nature, i.e. optimization problem.
Firstly, we have analyzed and reviewed existing video coding works from the perspective of optimization, and have concluded that block-based hybrid coding and end-to-end learned coding represent discrete and continuous optimization solutions in essence, respectively.
Secondly, based on the theoretical analysis about existing optimization, we have proposed a hybrid of discrete and continuous optimization video coding theory, which is more efficient, more refined, and more likely to achieve the globally optimal solution in theory.
Thirdly, guided by the theory, we have proposed a hybrid-optimization video coding framework based on deep networks entirely. We have performed extensive experiments to verify the efficiency of the proposed method. Results show that the proposed method outperforms the pure deep learned video coding methods and achieves comparable performance to HEVC reference software HM16.10 in PSNR.
More importantly, our hybrid optimization can serve as the bridge of both optimization to promote the improvement of coding performance together, so it has great potential.

There are several open issues for further study, some of which have been mentioned before. Firstly, we can add advanced schemes of the two frameworks into our framework to further realize the potential of hybrid optimization. Secondly, we need to find out better network structures for both high compression efficiency and low complexity. Thirdly, we can study the uni-directional inter-prediction for low-delay configuration.
\ifCLASSOPTIONcaptionsoff
   \newpage
\fi

\bibliographystyle{IEEEtran}
\bibliography{Video_Compression}




\end{document}